\documentclass[11pt,a4paper]{article}
\usepackage{jheppub}

\usepackage{lipsum}

\usepackage{verbatim}

\usepackage{graphicx} 
\usepackage{amsfonts,amsmath}
\usepackage{tikz, float}
\usepackage{braket}
\usepackage{subcaption}

\usepackage{import}
\usepackage{tikz}
\usepackage{tikz-3dplot}
\usetikzlibrary{shapes,decorations}
\usetikzlibrary{decorations.markings}
\usepackage{pgfplots}
\usepgfplotslibrary{fillbetween}
\usetikzlibrary{arrows.meta}
\pgfplotsset{
    compat=newest
}
\usetikzlibrary{shapes.geometric}
\usetikzlibrary{shapes.misc}
\tikzset{cross/.style={cross out, draw=black,thick, fill=none, minimum size=2*(#1-\pgflinewidth), inner sep=0pt, outer sep=0pt}, cross/.default={3pt}}
\usetikzlibrary {patterns.meta} \pgfdeclarepattern{
  name=hatch,
  parameters={\hatchsize,\hatchangle,\hatchlinewidth},
  bottom left={\pgfpoint{-.1pt}{-.1pt}},
  top right={\pgfpoint{\hatchsize+.1pt}{\hatchsize+.1pt}},
  tile size={\pgfpoint{\hatchsize}{\hatchsize}},
  tile transformation={\pgftransformrotate{\hatchangle}},
  code={
    \pgfsetlinewidth{\hatchlinewidth}
    \pgfpathmoveto{\pgfpoint{-.1pt}{-.1pt}}
    \pgfpathlineto{\pgfpoint{\hatchsize+.1pt}{\hatchsize+.1pt}}
    \pgfpathmoveto{\pgfpoint{-.1pt}{\hatchsize+.1pt}}
    \pgfpathlineto{\pgfpoint{\hatchsize+.1pt}{-.1pt}}
    \pgfusepath{stroke}
} }
\tikzset{
hatch size/.store in=\hatchsize,
hatch angle/.store in=\hatchangle,
hatch line width/.store in=\hatchlinewidth, hatch size=5pt,
hatch angle=0pt,
hatch line width=.5pt,
}
\usepgflibrary{shadings}

\usetikzlibrary{math}
\usetikzlibrary{calc}

\newcommand{\be}{\begin{equation}}
\newcommand{\ee}{\end{equation}}
\newcommand{\ba}{\begin{eqnarray}}
\newcommand{\ea}{\end{eqnarray}}

\newcommand{\AS}[1]{{\color{magenta}#1}}
\newcommand{\BM}[1]{{\color{blue}#1}}
\newcommand{\JDL}[1]{{\color{green}#1}}
\newcommand{\CS}[1]{{\color{orange}#1}}
\newcommand{\RED}[1]{{\color{red}#1}}

\newcommand{\lb}{\left(}
\newcommand{\rb}{\right)}

\newcommand{\vertiii}{\vert\kern-0.25ex\vert\kern-0.25ex\vert}

\title{Holography for BCFTs with Multiple Boundaries: Multi-Splitting Quenches}
\author[a,d]{Joseph Dominicus Lap}
\author[a,b]{Berndt M\"{u}ller}
\author[c]{Andreas Sch\"{a}fer}
\author[c]{Clemens Seidl}

\affiliation[a]{Department of Physics, Yale University,
New Haven, 06511, NY, USA}
\affiliation[b]{Department of Physics, Duke University, Durham, 27708, NC, USA}
\affiliation[c]{Institut f\"ur Theoretische Physik, Universit\"at Regensburg, Regensburg, D-93040, Germany}
\affiliation[d]{Instituut voor Theoretische Fysica K.U. Leuven, Celestijnenlaan 200D, B-3001, Leuven,  Belgium}

\emailAdd{Joseph.Dominicus.Lap@yale.edu}
\emailAdd{Berndt.Mueller@duke.edu}
\emailAdd{Andreas.Schaefer@physik.uni-regensburg.de}
\emailAdd{Clemens.Seidl@physik.uni-regensburg.de}

\abstract{We elaborate on the method introduced in arXiv:2403.02165 for holographic duals of Boundary Conformal Field Theories (BCFTs) with multiple boundaries. Using these advances we calculate the entanglement entropy as a function of time for 1+1-dimensional CFTs that are split into $N$ subsystems. We give explicit results for $N=4$ and $N=17$. We find that all qualitative differences that arise for larger $N$ 
are present $N=4$.}

\begin{document}

\maketitle

\section{Introduction}

Quantum Field Theories at criticality are well-described by Conformal Field Theories (CFTs). The symmetry of these systems is highly constraining, allowing analytic calculation of many quantities that would otherwise be intractable. Many systems of interest have boundaries (e.g. cells on a membrane \cite{Machta_2012}, spin-chains being joined \cite{Calabrese_2007}). However, typical replica-trick methods \cite{Calabrese:2004eu} become intractable with multiple boundaries as moduli become relevant. We extend the standard AdS/BCFT prescription \cite{Takayanagi:2011zk, Fujita_2011} to the case of multiple boundaries by using advances in the theory of conformal mapping and Riemann surfaces \cite{crowdy_solving_2020}. We then apply this prescription to the calculation of multiple splitting quenches viz. when a CFT on a line is split into $N=n+1$ pieces. The $n=1,2$ cases were studied previously \cite{Shimaji:2018czt,Caputa:2019avh}.

Why do we consider the case of a CFT splitting into multiple subsystems? In unitarily evolving systems entanglement dynamics are a rich probe of complicated physics. In black hole physics they have been used to heavily constrain black hole evolution \cite{PhysRevLett.44.301}, in trapped systems of cold atoms they are experimentally measurable \cite{Greiner_2002}, and it has been argued that it is key to understanding the hadronization process in heavy-ion collisions \cite{Muller:2022htn}. For strongly coupled-gauge theories dynamical observables are notoriously hard to calculate on a lattice \cite{Zhou_2022} and for systems with multiple boundaries the typical replica trick becomes non-trivial.  A 1+1D CFT on a domain that splits into multiple pieces at $t=0$ is thus both a non-trivial check of our formalism and interesting physics that should be experimentally realizable.


\section{Riemann Surfaces, Moduli Spaces, Uniformization}
Riemann surfaces are not standard material in mathematical methods in physics courses so we will provide a gentle introduction for physicists as suits our purposes (focusing on uniformization and conformal mappings rather than proving rigorous statements about holomorphic differentials etc.). For a more complete understanding of Riemann surfaces there are standard references \cite{farkas2012Riemann,fay1973theta} as well as more contemporary lectures \cite{eynard2018lecturesnotescompactRiemann,Dubrovin}.

\subsection{Moduli Spaces}
When working with a conformal field theory, our primary concern is with conformal invariance. For a simply connected subset of the complex plane \textendash\ say a disk or the upper half-plane \textendash\ the Riemann mapping theorem guarantees that all such subsets are conformally equivalent. However, we are concerned with worldsheets that are not simply connected. Consider an annulus in the complex plane with inner radius $r$ and outer radius $1$. 
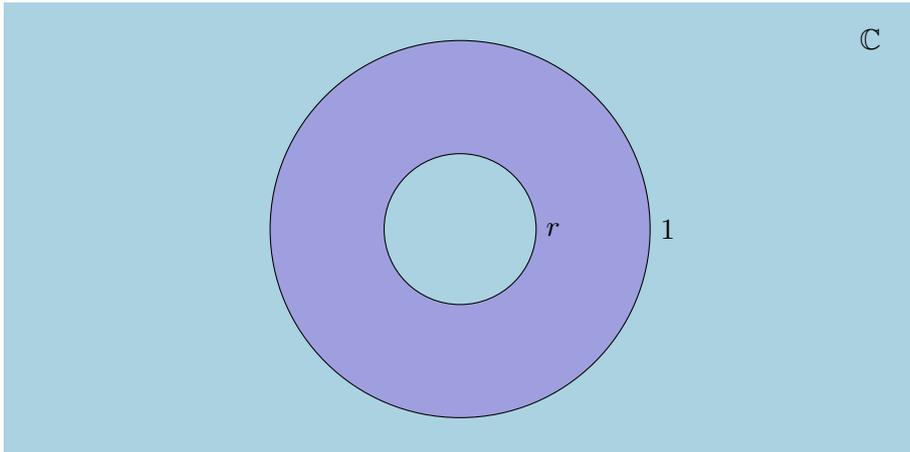
\begin{figure}[H]
\centering
\tikzmath{\l=3;\h=3;\b=1.5;\a=1;\r=1;}
\scalebox{1}{
\begin{tikzpicture}

    \path[fill=cyan!50!gray, opacity=0.5] ($(-2*\l,-\h)$)--($(2*\l,-\h)$)--($(2*\l,\h)$)--($(-2*\l,\h)$)--cycle;

    \node[above] at (1.8*\l,0.75*\h) {$\mathbb{C}$};

    \fill [white, even odd rule] (0,0) circle[radius=\r] circle[radius=2.5*\r];
    \fill [blue!50!gray, fill opacity=0.5, even odd rule, draw=black] (0,0) circle[radius=\r] circle[radius=2.5*\r];

    \node[right] at (\r,0) {$r$};
    \node[right] at (2.5*\r,0) {$1$};

\end{tikzpicture}
}
\caption{Annulus in the complex plane}
\end{figure}
Using the dilatation symmetry of the conformal group it is impossible to conformally map to another annulus with inner radius $r'\neq r$ and outer radius $1$. We can therefore parameterize the set of conformally inequivalent annuli with one real number: the ratio between the inner and outer radii. This is classically called a modulus, and spaces built out of such parameters are called moduli spaces. One such space that will be familiar to those who have taken a course in string theory is the moduli space of the torus, which is needed for the one-loop amplitude of the closed string. Since the torus can be constructed from a parallelogram with opposite sides identified, the typical coordinate is a complex number $\tau$ that specifies the top left corner.\footnote{One of course has to be more careful and specify to values of $\tau$ that uniquely identify tori. A quick exposition can be found in \cite{tong2012lecturesstringtheory} and a more in-depth one in standard string theory textbooks \cite{Polchinski_1998}.}

\begin{figure}[H]
\centering
\tikzmath{\l=3;\h=3;}
\scalebox{1}{
\begin{tikzpicture}

    \path[fill=cyan!50!gray, opacity=0.5] ($(-2*\l,-\h)$)--($(2*\l,-\h)$)--($(2*\l,\h)$)--($(-2*\l,\h)$)--cycle;

    \node[above] at (1.8*\l,0.75*\h) {$\mathbb{C}$};

    \def \a {\l};
    \def \b {0.5*\h};
    \def \t {0.5*\l};

    \path[fill=white] ($(-\a-0.5*\t,-\b)$)node[below left]{$0$}--($(-\a+0.5*\t,\b)$)node[above left]{$\tau$}--($(\a+0.5*\t,\b)$)node[above right]{$1+\tau$}--($(\a-0.5*\t,-\b)$)node[below right]{$1$} --cycle;

    \path[fill=blue!50!gray, fill opacity=0.5, draw=black, thick] ($(-\a-0.5*\t,-\b)$)
    --($(-\a+0.5*\t,\b)$)node[midway, sloped]{$\parallel$}
    --($(\a+0.5*\t,\b)$)node[midway, sloped]{$\vertiii$}
    --($(\a-0.5*\t,-\b)$)node[midway, sloped]{$\parallel$}
    --cycle node[midway, sloped]{$\vertiii$}
    ;

\end{tikzpicture}
}
\caption{Modulus of a torus, opposite sides are identified.}

\end{figure}
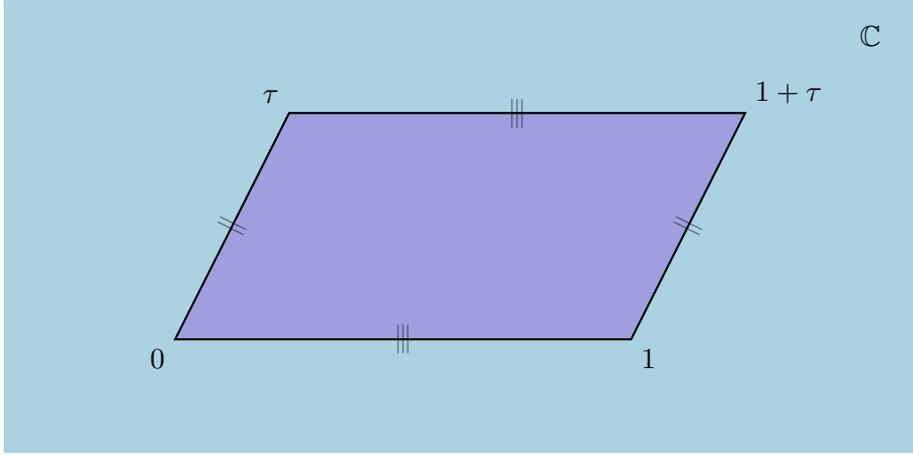

Generally, to imbue these spaces with nice features and to prove facts about them is difficult, but to say how large they are is simple. In fact, Riemann calculated that compact Riemann surfaces of genus $g>1$ are characterized by $6g-6$ real moduli \cite{Riemann1857}. One might na\"ively worry that as we increase the number of boundaries we will end up with an unwieldy high-dimensional moduli space to contend with. In many instances this is not the case as one can just work in a specific part of the space – in appendix \ref{sec:appendix_SKP_asymptotics} we show that for splitting quenches this is the case.

\subsection{Riemann Surfaces}
We are concerned in this paper with worldsheets that are not simply connected, {\it viz.}, with multiple line segments or disks cut out of them. We wish then to conformally map them to a uniformized space where calculations of physical observables are tractable. Conformal maps do not change topology\footnote{Biholomorphisms are homeomorphisms.} and therefore for a given worldsheet we need only consider surfaces with the same topological invariants. The two relevant invariants here are the number of boundaries and the genus of the surface (since we are working with subsets of the plane this will always be 0). These are the $g=0$ Riemann surfaces with $n$ boundaries and one marked point, the point at infinity which we add to compactify the plane). The full compactified moduli space of such surfaces is complicated and may not even be orientable \cite{Pandharipande_2024}. However, for our purposes (uniformizing the Riemann surfaces so we can do calculations) these difficulties do not enter. There is a canonical doubling of Riemann surfaces with $n$ boundaries to compact Riemann surfaces of genus $g=n-1$ (the Schottky doubling). The original surface with boundary can then be obtained by an involution on the doubling.\footnote{See section II.1.3 of \cite{ahlfors2015riemann} for an in-depth discussion.}
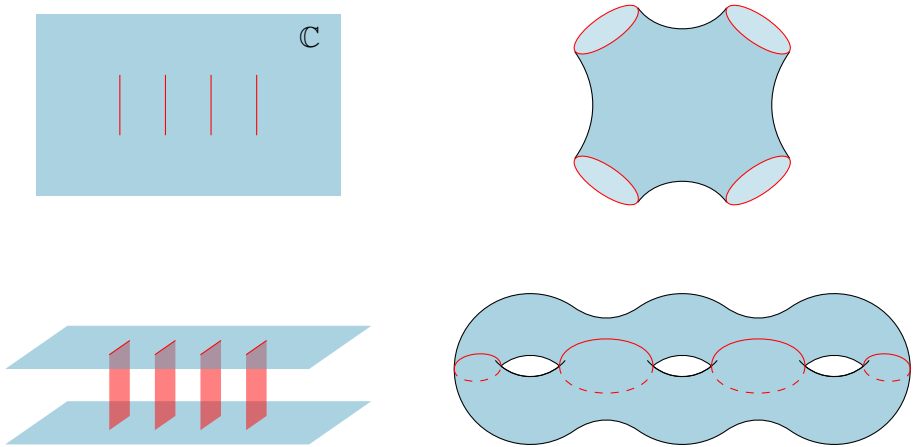
\begin{figure}[H]
\centering
\tikzmath{\l=6.5;\h=3.5;}
\scalebox{1}{
\begin{tikzpicture}

    \coordinate (al) at (0,0);
    \coordinate (ar) at (\l,0);
    \coordinate (bl) at (0,-\h);
    \coordinate (br) at (\l,-\h);

    \def \a {1};  
    \def \rot {35};
    \def \ew {1};
    \def \eh {0.3};
    
    \node[ellipse, rotate=\rot, minimum width=\ew cm, minimum height=\eh cm] (aral) at ($(ar) + (-\a,\a)$) {};
    \node[ellipse, rotate=-\rot, minimum width=\ew cm, minimum height=\eh cm] (arar) at ($(ar) + (\a,\a)$) {};
    \node[ellipse, rotate=-\rot, minimum width=\ew cm, minimum height=\eh cm] (arbl) at ($(ar) + (-\a,-\a)$) {};
    \node[ellipse, rotate=\rot, minimum width=\ew cm, minimum height=\eh cm] (arbr) at ($(ar) + (\a,-\a)$) {};

    \path[fill=cyan!50!gray, semitransparent]
    (aral.east) to [out=\rot-90,in=180-\rot+90] (arar.west)
    -- (arar.east) to [out=-\rot-90,in=\rot+90] (arbr.east)
    -- (arbr.west) to [in=-\rot+90,out=\rot+90] (arbl.east)
    -- (arbl.west) to [in=\rot-90,out=-\rot+90] (aral.west)
    -- cycle
    ;

    \node[ellipse, fill=white, rotate=\rot, minimum width=\ew cm, minimum height=\eh cm] (aral) at ($(ar) + (-\a,\a)$) {};
    \node[ellipse, fill=white, rotate=-\rot, minimum width=\ew cm, minimum height=\eh cm] (arar) at ($(ar) + (\a,\a)$) {};
    \node[ellipse, fill=white, rotate=-\rot, minimum width=\ew cm, minimum height=\eh cm] (arbl) at ($(ar) + (-\a,-\a)$) {};
    \node[ellipse, fill=white, rotate=\rot, minimum width=\ew cm, minimum height=\eh cm] (arbr) at ($(ar) + (\a,-\a)$) {};

    \draw[] (aral.east) to [out=\rot-90,in=180-\rot+90] (arar.west);
    \draw[] (arbl.east) to [out=-\rot+90,in=\rot+90] (arbr.west);
    \draw[] (aral.west) to [out=\rot-90,in=-\rot+90] (arbl.west);
    \draw[] (arar.east) to [out=-\rot-90,in=\rot+90] (arbr.east);

    \node[ellipse, draw=red, fill=cyan!50!gray, fill opacity=0.3, rotate=\rot, minimum width=\ew cm, minimum height=\eh cm] (aral) at ($(ar) + (-\a,\a)$) {};
    \node[ellipse, draw=red, fill=cyan!50!gray, fill opacity=0.3, rotate=-\rot, minimum width=\ew cm, minimum height=\eh cm] (arar) at ($(ar) + (\a,\a)$) {};
    \node[ellipse, draw=red, fill=cyan!50!gray, fill opacity=0.3, rotate=-\rot, minimum width=\ew cm, minimum height=\eh cm] (arbl) at ($(ar) + (-\a,-\a)$) {};
    \node[ellipse, draw=red, fill=cyan!50!gray, fill opacity=0.3, rotate=\rot, minimum width=\ew cm, minimum height=\eh cm] (arbr) at ($(ar) + (\a,-\a)$) {};


    \def \a {1};
    \def \ang {55};

    \draw[fill=cyan!50!gray, fill opacity=0.5] ($(br) + (-3*\a, 0)$) arc (180:\ang:\a)
    to [out=\ang-90,in=-90-\ang]
    ($(br) + (180-\ang:\a)$) arc (180-\ang:\ang:\a)
    to [out=\ang-90,in=-90-\ang]
    ($(br) + (2*\a, 0) + (180-\ang:\a)$) arc (180-\ang:\ang-180:\a)
    to [out=\ang+90,in=90-\ang]
    ($(br) + (-\ang:\a)$) arc (-\ang:-180+\ang:\a)
    to [out=\ang+90,in=90-\ang]
    ($(br) + (-2*\a, 0) + (-\ang:\a)$) arc (-\ang:-180:\a)
    ;

    \def \b {0.6*\a};
    \draw[fill=white] ($(br) + (0, 0.5*\a) + (-130:\b)$)
    arc (-130:-50:\b)
    arc (50:130:\b)
    ;
    \draw[] ($(br) + (0, 0.5*\a) + (-140:\b)$)
    arc (-140:-40:\b)
    ;

    \draw[fill=white] ($(br) + (-2*\a, 0.5*\a) + (-130:\b)$)
    arc (-130:-50:\b)
    arc (50:130:\b)
    ;
    \draw[] ($(br) + (-2*\a, 0.5*\a) + (-140:\b)$)
    arc (-140:-40:\b)
    ;

    \draw[fill=white] ($(br) + (2*\a, 0.5*\a) + (-130:\b)$)
    arc (-130:-50:\b)
    arc (50:130:\b)
    ;
    \draw[] ($(br) + (2*\a, 0.5*\a) + (-140:\b)$)
    arc (-140:-40:\b)
    ;

    \draw[red,dashed] ($(br) + (-3*\a, 0)$)
    to [out=-90,in=-90]
    ($(br) + (-2*\a, 0.5*\a) + (-130:\b)$);
    \draw[red] ($(br) + (-3*\a, 0)$)
    to [out=90,in=90]
    ($(br) + (-2*\a, 0.5*\a) + (-130:\b)$);

    \draw[red,dashed] ($(br) + (-2*\a, 0.5*\a) + (-50:\b)$)
    to [out=-90,in=-90]
    ($(br) + (0, 0.5*\a) + (-130:\b)$);
    \draw[red] ($(br) + (-2*\a, 0.5*\a) + (-50:\b)$)
    to [out=90,in=90]
   ($(br) + (0, 0.5*\a) + (-130:\b)$);

    \draw[red,dashed] ($(br) + (2*\a, 0.5*\a) + (-130:\b)$)
    to [out=-90,in=-90]
    ($(br) + (0, 0.5*\a) + (-50:\b)$);
    \draw[red] ($(br) + (2*\a, 0.5*\a) + (-130:\b)$)
    to [out=90,in=90]
   ($(br) + (0, 0.5*\a) + (-50:\b)$);

    \draw[red,dashed] ($(br) + (3*\a, 0)$)
    to [out=-90,in=-90]
    ($(br) + (2*\a, 0.5*\a) + (-50:\b)$);
    \draw[red] ($(br) + (3*\a, 0)$)
    to [out=90,in=90]
    ($(br) + (2*\a, 0.5*\a) + (-50:\b)$);


    \def \cl {2};
    \def \ch {1.2};

    \path[fill=cyan!50!gray, semitransparent] ($(al)+(-\cl,-\ch)$) --($(al)+(-\cl,\ch)$) --($(al)+(\cl,\ch)$) --($(al)+(\cl,-\ch)$) --cycle;

    \def \a {0.4};
    \def \b {0.3};
    \draw[red] (-3*\b,-\a) -- (-3*\b,\a);
    \draw[red] (-\b,-\a) -- (-\b,\a);
    \draw[red] (\b,-\a) -- (\b,\a);
    \draw[red] (3*\b,-\a) -- (3*\b,\a);

    \node[] at (0.8*\cl,0.75*\ch) {$\mathbb{C}$};


    \def \d {1};
    \def \a {2};
    \def \b {1};
    \def \ang {35};
    \def \cut {0.3};

    \coordinate (bl) at ({-0.5*cos(\ang)},-\h);  

    \path[fill=cyan!50!gray, fill opacity=0.5]
    ($(bl)+(-\a,-\d)$)
    --($(bl)+(-\a, -\d)+(\ang:\b)$)
    --($(bl)+(\a, -\d)+(\ang:\b)$)
    --($(bl)+(\a, -\d)$)--cycle
    ;

    \path[fill=red, fill opacity=0.5]
    ($(bl)+(-3*\cut,-\d)+(\ang:0.33*\b)$)
    --($(bl)+(-3*\cut, 0)+(\ang:0.33*\b)$)
    --($(bl)+(-3*\cut, 0)+(\ang:0.66*\b)$)
    --($(bl)+(-3*\cut, -\d)+(\ang:0.66*\b)$)--cycle
    ;

    \path[fill=red, fill opacity=0.5]
    ($(bl)+(-\cut,-\d)+(\ang:0.33*\b)$)
    --($(bl)+(-\cut, 0)+(\ang:0.33*\b)$)
    --($(bl)+(-\cut, 0)+(\ang:0.66*\b)$)
    --($(bl)+(-\cut, -\d)+(\ang:0.66*\b)$)--cycle
    ;

    \path[fill=red, fill opacity=0.5]
    ($(bl)+(\cut,-\d)+(\ang:0.33*\b)$)
    --($(bl)+(\cut, 0)+(\ang:0.33*\b)$)
    --($(bl)+(\cut, 0)+(\ang:0.66*\b)$)
    --($(bl)+(\cut, -\d)+(\ang:0.66*\b)$)--cycle
    ;

    \path[fill=red, fill opacity=0.5]
    ($(bl)+(3*\cut,-\d)+(\ang:0.33*\b)$)
    --($(bl)+(3*\cut, 0)+(\ang:0.33*\b)$)
    --($(bl)+(3*\cut, 0)+(\ang:0.66*\b)$)
    --($(bl)+(3*\cut, -\d)+(\ang:0.66*\b)$)--cycle
    ;

    \path[fill=cyan!50!gray, fill opacity=0.5]
    ($(bl)+(-\a,0)$)
    --($(bl)+(-\a, 0)+(\ang:\b)$)
    --($(bl)+(\a, 0)+(\ang:\b)$)
    --($(bl)+(\a, 0)$)--cycle
    ;

    \draw[red] ($(bl)+(-3*\cut, 0)+(\ang:0.33*\b)$)
    --($(bl)+(-3*\cut, 0)+(\ang:0.66*\b)$);
    \draw[red] ($(bl)+(-\cut, 0)+(\ang:0.33*\b)$)
    --($(bl)+(-\cut, 0)+(\ang:0.66*\b)$);
    \draw[red] ($(bl)+(\cut, 0)+(\ang:0.33*\b)$)
    --($(bl)+(\cut, 0)+(\ang:0.66*\b)$);
    \draw[red] ($(bl)+(3*\cut, 0)+(\ang:0.33*\b)$)
    --($(bl)+(3*\cut, 0)+(\ang:0.66*\b)$);

\end{tikzpicture}
}
\caption{Schottky doubling of the $g=0$, 4-boundaried Riemann surface}
\label{fig:doubling}
\end{figure}

For simplicity's sake we work with compact Riemann surfaces.\footnote{There is a subtlety here in that how one glues the doubling is defined up to a twist. Physically one can think of this as boundary conditions enforced on the boundaries in order to maintain conformal invariance. These are in general not unique. The effect of this in holography is a tension term which affects the extension of the boundary into the bulk \cite{Fujita_2011}. However, the only effect on geodesic structure is the addition of a constant to the disconnected geodesic length. Therefore the time at which the saddle-points switch is dependent on this choice. We will choose the trivial option of no twist and no boundary entropy and plot both curves.} Every compact Riemann surface is representable as a complex algebraic curve, and therefore the moduli space at hand is the same as the extremely well studied moduli space of algebraic curves \cite{Deligne1969}. Although this space is notoriously subtle, for all the examples we know of these subtleties are irrelevant to the calculations at hand.

Our focus in this paper will be a CFT on a line that splits into $N=n+1$ pieces, where the appropriate worldsheet is the complex plane minus $n$ parallel line segments of the same ``small" length $2a$, i.e., where $a$ is much smaller than the distance between any two of the line segments (we show this in section \ref{sec:BCFT_and_splitting_quenches}). The Schottky doubling makes clear that our worldsheet is the projection onto $\mathbb{C}$ of the standard branched cover of $\mathbb{C}$ representation of a hyperelliptic curve (see figure~\ref{fig:doubling}).
The fact that $a$ is small allows us to use asymptotics of the Schottky-Klein prime function and therefore to have an explicit conformal map. The remaining moduli that matter are the relative placements of the slits in the complex plane. We can then calculate the periods of the $A$ and $B$ cycles of the worldsheet (which is needed for the black hole entropy term) using the Abel-Jacobi map\footnote{This amounts to doing a hyperelliptic integral which is easily done numerically for a given example. In \cite{Lap:2024hsy} we used the Abel-Jacobi map as a conformal map; this cannot be done for $g\neq1$. The Abel-Jacobi map has as its image the Jacobi torus of the curve $T^{2g}$ – the "space of cycles" broadly speaking. For $g=1$ we are mapping a torus to a torus and this is an isomorphism, but for $g\neq1$ one cannot rely on this fact.} and uniformize with the standard Schottky uniformization.
\subsection{Uniformization}
The ``uniformization'' of a Riemann surface is broadly speaking a way of representing it such that its dependence on its moduli is clear. Mathematicians in the $19^{\rm th}$ century developed many notions and methods of uniformization, but the one most contemporary readers will be familiar with is ``Fuchsian uniformization" in which one represents a given Riemannn surface $R$ as a quotient of the upper half-plane\footnote{$\mathbb{H}$ denotes the upper half-plane with hyperbolic metric. Where $\mathbb{H}_n$ is often referred to as the poincar\'e patch of Euclidean $AdS_n$ in physics literature or the poincar\'e half-plane model in the mathematics literature. The metric is the standard: $ds^2=\frac{dz^2+\sum_{i=1}^{n-1} dx_i^2}{z^2}$} $\mathbb{H}$ (or equivalently its fundamental polygon) by a subgroup $\Gamma\subset {\rm PSL}_2(\mathbb{R})$: $R\sim \mathbb{H}/\Gamma$. $\Gamma$ is then isomorphic to the fundamental group.

A less well-known method of uniformization is ``Schottky uniformization" in which one represents a Riemann surface $R$ as a quotient of a not simply-connected domain $\Omega$ (the unit disk with multiple disks cut out) by a subgroup $\Theta \subset {\rm PSL}_2(\mathbb{C})$: $R\sim\Omega/\Theta$. The benefit of uniformizing the Riemann surface in this way is that it makes holography simple. The isomorphism group of $\mathbb{H}_3$ is ${\rm SL}_2(\mathbb{C})$, and therefore the quotient on the boundary extends into the bulk and defines a 3-manifold (see \cite{Krasnov:2000zq,Carlip_1995} for in-depth discussions). Furthermore, the conformal map to the fundamental domain can be constructed explicitly, and therefore the way the bulk transforms is known \cite{Roberts:2012aq} (see section~\ref{sec:Holography_for_Multiple_Boundaries}).

\begin{figure}[H]
\centering
\tikzmath{\l=6;\h=3.5;}
\scalebox{1}{
\begin{tikzpicture}

    \coordinate (al) at (0,0);
    \coordinate (ar) at (\l,0);
    \coordinate (bl) at (0,-\h);
    \coordinate (br) at (\l,-\h);


    \def \a {1};
    \def \ang {55};

    \draw[fill=cyan!50!gray, fill opacity=0.5] ($(al) + (-2*\a, 0)$) arc (180:\ang:\a)
    to [out=\ang-90,in=-90-\ang]
    ($(al) + (\a, 0) + (180-\ang:\a)$) arc (180-\ang:\ang-180:\a)
    to [out=\ang+90,in=90-\ang]
    ($(al) + (-\a, 0) + (-\ang:\a)$) arc (-\ang:-180:\a)
    ;

    \def \b {0.6*\a};
    
    \draw[fill=white] ($(al) + (-\a, 0.5*\a) + (-130:\b)$)
    arc (-130:-50:\b)
    arc (50:130:\b)
    ;
    \draw[] ($(al) + (-\a, 0.5*\a) + (-140:\b)$)
    arc (-140:-40:\b)
    ;

    \draw[fill=white] ($(al) + (\a, 0.5*\a) + (-130:\b)$)
    arc (-130:-50:\b)
    arc (50:130:\b)
    ;
    \draw[] ($(al) + (\a, 0.5*\a) + (-140:\b)$)
    arc (-140:-40:\b)
    ;

    \def \angg {60};

    \begin{scope}[thick, decoration={
        markings,
        mark=at position 0.5 with {\arrow{<}}}
        ]

        \draw[green, postaction={decorate}] (al)
        to [out=10,in=180-20]
        ($(al) + (\a, 0.5*\a) + (-110:\b)$)
        ;
        \draw[green, dotted] ($(al) + (\a, 0.5*\a) + (-110:\b)$)
        to [out=-20,in=-10]
        ($(al) + (\a, 0) + (-100:\a)$)
        ;
        \draw[green, postaction={decorate}] ($(al) + (\a, 0) + (-100:\a)$)
        to [out=180-10,in=-60]node[pos=0, below]{$D$}
        (al)
        ;

        \draw[orange, postaction={decorate}] (al)
        to [out=180-10,in=20]
        ($(al) + (-\a, 0.5*\a) + (-70:\b)$)
        ;
        \draw[orange, dotted] ($(al) + (-\a, 0.5*\a) + (-70:\b)$)
        to [out=180+20,in=180+10]
        ($(al) + (-\a, 0) + (-80:\a)$)
        ;
        \draw[orange, postaction={decorate}] ($(al) + (-\a, 0) + (-80:\a)$)
        to [out=10,in=180+60]node[pos=0, below]{$A$}
        (al)
        ;
    \end{scope}

    \begin{scope}[thick,decoration={
        markings,
        mark=at position 0.5 with {\arrow{>}}}
        ]
        
        \draw[red, postaction={decorate}] (al)
        to [out=\angg,in=90]node[pos=0.6, above]{$C$}
        ($(al) + (1.75*\a, 0)$)
        ;
        \draw[red, postaction={decorate}] ($(al) + (1.75*\a, 0)$)
        to [out=-90,in=-\angg]
        (al)
        ;

        \draw[blue, postaction={decorate}] (al)
        to [out=180-\angg,in=90]node[pos=0.6, above]{$B$}
        ($(al) + (-1.75*\a, 0)$)
        ;
        \draw[blue, postaction={decorate}] ($(al) + (-1.75*\a, 0)$)
        to [out=-90,in=180+\angg]
        (al)
        ;
    \end{scope}


    \def \r {1.25};
    \def \ang {45};

    \path[fill=cyan!50!gray, fill opacity=0.5]
    ($(bl)+(0.5*\ang:\r)$) -- ($(bl)+(1.5*\ang:\r)$) -- ($(bl)+(2.5*\ang:\r)$) -- ($(bl)+(3.5*\ang:\r)$) -- ($(bl)+(4.5*\ang:\r)$) -- ($(bl)+(5.5*\ang:\r)$) -- ($(bl)+(6.5*\ang:\r)$) -- ($(bl)+(7.5*\ang:\r)$) -- cycle;

    \begin{scope}[thick,decoration={
        markings,
        mark=at position 0.5 with {\arrow{>}}}
        ]
        
        \draw[orange, postaction={decorate}] 
        ($(bl)+(5.5*\ang:\r)$) -- ($(bl)+(6.5*\ang:\r)$)node[pos=0.5, below]{$A$}
        ;
        \draw[orange, postaction={decorate}]
        ($(bl)+(7.5*\ang:\r)$) -- ($(bl)+(0.5*\ang:\r)$)node[pos=0.5, right]{$A^{-1}$}
        ;

        \draw[blue, postaction={decorate}] 
        ($(bl)+(6.5*\ang:\r)$) -- ($(bl)+(7.5*\ang:\r)$)node[pos=0.5, below right]{$B$}
        ;
        \draw[blue, postaction={decorate}]
        ($(bl)+(0.5*\ang:\r)$) -- ($(bl)+(1.5*\ang:\r)$)node[pos=0.5, above right]{$B^{-1}$}
        ;

        \draw[red, postaction={decorate}] 
        ($(bl)+(1.5*\ang:\r)$) -- ($(bl)+(2.5*\ang:\r)$)node[pos=0.5, above]{$C$}
        ;
        \draw[red, postaction={decorate}]
        ($(bl)+(3.5*\ang:\r)$) -- ($(bl)+(4.5*\ang:\r)$)node[pos=0.5, left]{$C^{-1}$}
        ;

        \draw[green, postaction={decorate}] 
        ($(bl)+(2.5*\ang:\r)$) -- ($(bl)+(3.5*\ang:\r)$)node[pos=0.5, above left]{$D$}
        ;
        \draw[green, postaction={decorate}]
        ($(bl)+(4.5*\ang:\r)$) -- ($(bl)+(5.5*\ang:\r)$)node[pos=0.5, below left]{$D^{-1}$}
        ;

    \end{scope}


    \def \r {1.25};
    \def \re {0.25*\r}

    \draw[dotted, thick, blue]
    ($(ar) + (0:\r)$) arc (0:180:{\r} and {\re})
    ;
    \draw[fill=cyan!50!gray, fill opacity=0.5]
    ($(ar) + (0:\r)$) arc (0:360:\r)
    ;
    \draw[blue, thick]
    ($(ar) + (0:\r)$) arc (0:-180:{\r} and {\re})
    ;

    \def \ew {0.1};
    \def \eh {0.2};

    \coordinate (e1) at ($(ar)+(0.8*\r,0.4*\r)$);
    \coordinate (e1p) at ($(ar)+(0.7*\r,-0.3*\r)$);

    \draw[red, thick, dotted]
    (e1) arc (90:-90:{\ew} and {\eh})
    ;
    \draw[red, thick, dotted]
    (e1p) arc (90:-90:{\ew} and {\eh})
    ;

    \draw[fill=red!50!gray, fill opacity=0.5]
    (e1) arc (90:270:{\ew} and {\eh})
    .. controls ($(ar)+(1.25*\r,0.1*\r)$) and ($(ar)+(1.25*\r,-0.4*\r)$)  ..
    (e1p)
    arc (90:270:{\ew} and {\eh})
    .. controls ($(ar)+(1.7*\r,-0.65*\r)$) and ($(ar)+(1.7*\r,0.4*\r)$)  ..
    (e1)
    ;
    
    \draw[red, thick]
    (e1) arc (90:270:{\ew} and {\eh})
    ;
    \draw[red, thick]
    (e1p) arc (90:270:{\ew} and {\eh})
    ;

    \def \ew {0.1};
    \def \eh {0.2};

    \coordinate (e1) at ($(ar)+(-0.6*\r,0.6*\r)$);
    \coordinate (e1p) at ($(ar)+(-0.5*\r,-0.4*\r)$);

    \draw[orange, thick, dotted]
    (e1) arc (90:270:{\ew} and {\eh})
    ;
    \draw[orange, thick, dotted]
    (e1p) arc (90:270:{\ew} and {\eh})
    ;

    \draw[fill=orange!50!gray, fill opacity=0.5]
    (e1) arc (90:-90:{\ew} and {\eh})
    .. controls ($(ar)+(-1.3*\r,0.4*\r)$) and ($(ar)+(-1.25*\r,-0.6*\r)$)  ..
    (e1p)
    arc (90:-90:{\ew} and {\eh})
    .. controls ($(ar)+(-1.8*\r,-0.95*\r)$) and ($(ar)+(-1.6*\r,0.8*\r)$)  ..
    (e1)
    ;
    
    \draw[orange, thick]
    (e1) arc (90:-90:{\ew} and {\eh})
    ;
    \draw[orange, thick]
    (e1p) arc (90:-90:{\ew} and {\eh})
    ;


    \def \r {1.25};

    \draw[fill=cyan!50!gray, fill opacity=0.5, draw=blue, thick]
    ($(br) + (0:\r)$) arc (0:360:\r)
    ;

    \coordinate (c1) at ($(br)+(0.4*\r,-0.4*\r)$);
    \coordinate (c2) at ($(br)+(-0.5*\r,0.4*\r)$);

    \def \re {0.2*\r};
    \draw[fill=white, draw=red, thick]
    ($(c1) + (0:\re)$) arc (0:360:\re)
    ;
    \def \re {0.2*\r};
    \draw[fill=white, draw=orange, thick]
    ($(c2) + (0:\re)$) arc (0:360:\re)
    ;

\end{tikzpicture}
}
\caption{Fuchsian uniformization (left) and Schottky uniformization (right) of the genus 2 surface. On the left the fundamental domain is simply connected and is conformally equivalent to the upper half-plane by Schwarz-Christoffel mapping. The genus 2 surface is obtained by identifying sides of the fundamental polygon and inversely the fundamental polygon is obtained by cutting the surface along cycles. On the right the fundamental domain is not simply connected. The genus 2 surface is obtained by reflecting across the unit disk and identifying the interior disks with their reflection. The fundamental region can be recovered via an involution on the surface.
}
\end{figure}
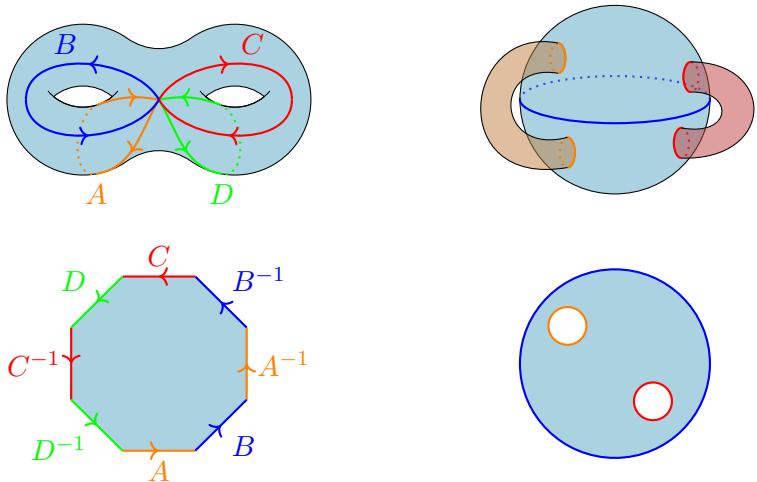

Recent work by Crowdy has greatly simplified the construction of conformal maps from multiply connected domains (equivalently, Riemann surfaces with boundaries) to their Schottky uniformization \cite{crowdy_solving_2020,crowdy_conformal_2012,crowdy_schottky-klein_2011}. These conformal maps are built off of a special function associated with the disconnected domain of a given Schottky uniformization and is known as the Schottky-Klein prime function. It has a well-known infinite product representation \cite{baker_abelian_nodate},
\be 
\omega(z,\alpha)=(z-\alpha)\prod_{\theta_k\in\Tilde{\Theta}}\frac{( z-\theta_k(\alpha))(\alpha-\theta_k(z))}{(\alpha-\theta_k(\alpha))(z-\theta_k(z))}\;,\label{eq:SKP_function}
\ee
where $\tilde{\Theta}$ includes all elements of $\Theta$ except inverses and the identity, and $\alpha$ is a chosen point in $\Omega$.

Then, conformal maps can be built from the domain $\Omega$ to the desired multiply connected domain from the Schottky-Klein prime function. For example, for the case of parallel line segments that are perpendicular to the real axis (the worldsheet for multiple splitting quenches, see section~\ref{sec:BCFT_and_splitting_quenches}) the appropriate conformal map is
\be
\phi(z)=\frac{1}{2\bar{\alpha}}+\frac{1}{2\alpha}-\partial_\alpha \ln(\omega(z;\alpha))-\frac{1}{\bar{\alpha}^2}\partial_{\bar{\alpha}^{-1}}\ln(\omega(z;\bar{\alpha}^{-1}))\;.
\ee
Here $\alpha$ has the additional significance of being the pre-image of the point at infinity.\footnote{We add this point to our worldsheet for the conformal compactification as is standard, but remove it after the conformal mapping from the worldsheet and $\Omega$. Otherwise interactions can occur between $x=-\infty$ and $x=\infty$ which is unphysical. Holographically this is treated by excluding the geodesics that are not homologous to the boundary due to the missing point.}

There are algorithms for calculating the Schottky-Klein prime function (and therefore the appropriate conformal maps) but they get increasingly slow as the genus grows and must be calculated for specific moduli \cite{Crowdy2007}.
However, in the case of a splitting quench the boundary is introduced as a regulator and taken to 0. We use the regulator length as a small parameter to calculate an asymptotic expression for the Schottky-Klein prime function in appendix~\ref{sec:appendix_SKP_asymptotics}. This makes analytic calculation of the conformal map possible regardless of the number of boundaries. We would like to emphasize that this is a general method applicable to any multiply connected domain where the moduli are small.

\section{Holography for Multiple Boundaries}\label{sec:Holography_for_Multiple_Boundaries}

Consider a CFT on a 2D worldsheet in the vacuum state whose holographic dual is empty Euclidean AdS$_3$ with the  Poincar\'{e} metric
\begin{align}
\label{poincare}ds^2=\frac{1}{z^2}\left(dz^2+dw d\bar{w}\right) ,
\end{align}
where $w=x+it,\bar{w}=x-it$ are complex coordinates,
the curvature of AdS is as usual set to one, and the CFT lives on the conformal boundary of the AdS space at $z\rightarrow0$.
What happens if the CFT has boundaries? Generally, boundaries break conformal invariance, but it is possible to choose boundary conditions, so-called conformal boundary conditions, such that the conformal group is broken to a smaller one \cite{cardy2008boundaryconformalfieldtheory}. Taking a bottom-up approach \cite{Fujita_2011}, we can add the Gibbons-Hawking term to the Einstein-Hilbert action and let the gravitational path integral tell us how the boundary extends into the bulk. For line-segments this boundary can be quite complicated (see figure 12 of \cite{Shimaji:2018czt}) and it can vary in a non-trivial way as one changes the size of the boundary. This makes it highly non-trivial to find expressions for holographic observables that depend on the boundary length. The solution is to use the conformal invariance to map the BCFT to its Schottky uniformization where the calculation of observables simplifies dramatically.

In order to make use of the uniformization procedure presented in the preceding section one has to understand the transformation of the bulk AdS space induced by a conformal transformation $w=f(\eta)$ of the CFT attached to it.
The conformal transformation on the boundary is equivalent to the following coordinate transformation in the AdS$_3$ bulk \cite{Roberts:2012aq}
\begin{align}
\begin{split}
    w&=f(\eta)-\frac{2\zeta^2f'^2\bar{f}''}{4|f'|^2+\zeta^2|f''|^2}\\
    \bar{w}&=\bar{f}(\bar{\eta})-\frac{2\zeta^2\bar{f}'^2f''}{4|f'|^2+\zeta^2|f''|^2}\\
    z&=\frac{4\zeta (f'\bar{f}')^{3/2}}{4|f'|^2+\zeta^2|f''|^2}\label{eq:holo_dual_bulk_transform}\;,
\end{split}
\end{align}
where $(\eta,\Bar{\eta},\zeta)$ denotes a different set of bulk coordinates.

The most general solution to Einstein's equations with negative cosmological constant in three dimensions is the metric \cite{Banados:1992wn}
\begin{align}
ds^2=\frac{d\zeta^2}{\zeta^2}+L(\eta)d\eta^2+\bar{L}(\bar{\eta})d\bar{\eta}^2+\left(\frac{\zeta^2}{2}+2L(\eta)\bar{L}(\bar{\eta})\zeta^{-2}\right)d\eta d\bar{\eta}\;,\label{eq:splitting_banados_metric}
\end{align}
where $\eta=\Xi+i\Upsilon,\bar{\eta}=\Xi-i\Upsilon$ are complex null coordinates of the theory living on the boundary of AdS and $L,\bar{L}$ are arbitrary functions of $\eta,\bar{\eta}$.
Note that the metric is only globally defined for constant $L$ and $\bar{L}$. 

Balasubramanian and Kraus \cite{Balasubramanian:1999re} showed that the holomorphic and anti-holomorphic stress tensors of asymptotically local AdS spacetimes may be defined in terms of these functions, i.e.,
\begin{align}
    T(\eta)=-\frac{1}{16\pi }L(\eta)\;,\hspace{1cm} \bar{T}(\bar{\eta})=-\frac{1}{16\pi }\bar{L}(\bar{\eta})\;.
\end{align}
The stress tensor transforms as expected under conformal transformations,
\begin{align}
T(\eta)\rightarrow f'(\eta)T(f(\eta))+\frac{c}{12}\{f(\eta),\eta\}\;,
\end{align}
where $c$ is the central charge of the CFT and $\{\cdot,\cdot\}$ denotes the Schwarzian derivative $\{f(x),x\}=\frac{3f''^2-2f'''f'}{4f'^2}$, which can be thought of as a measure of the degree to which a given conformal map differs from a M\"obius map. Provided that $f(\eta)$ is the conformal map to the vacuum solution of AdS, the above definition of the stress tensor reduces to just the Schwarzian derivative, implying 
\begin{align}
L(\eta)\propto \{f(\eta),\eta\}\;,\hspace{1cm}\bar{L}(\bar{\eta})\propto\{\bar{f}(\bar{\eta}),\bar{\eta}\}\;.
\end{align}
Therefore, in order to find the bulk metric one only needs to calculate the Schwarzian derivative of the conformal map from the setup to the vacuum. Our plan will then be as follows:
\begin{enumerate}
    \item Find the appropriate worldsheet;
    \item Find the conformal map from the Schottky uniformization to the worldsheet;\footnote{Constructing conformal maps between multiply connected domains is in general hard. See \cite{crowdy_solving_2020} for a selection of maps for common domains in terms of the Schottky-Klein prime function. For small moduli one can apply the asymptotics calculated in appendix~\ref{sec:appendix_SKP_asymptotics}.}
    \item Calculate everything holographically in the Schottky double of the fundamental domain;
    
    \item Express the result in the original physical coordinates.
\end{enumerate}
In the following section we apply this plan to the problem of finding the real-time evolution of entanglement entropy of spatial subsystems for a 1+1D CFT that splits into multiple pieces at $t=0$.

\section{Holographic entanglement entropy for multi-splitting quenches}
\subsection{BCFT and splitting quenches}\label{sec:BCFT_and_splitting_quenches}

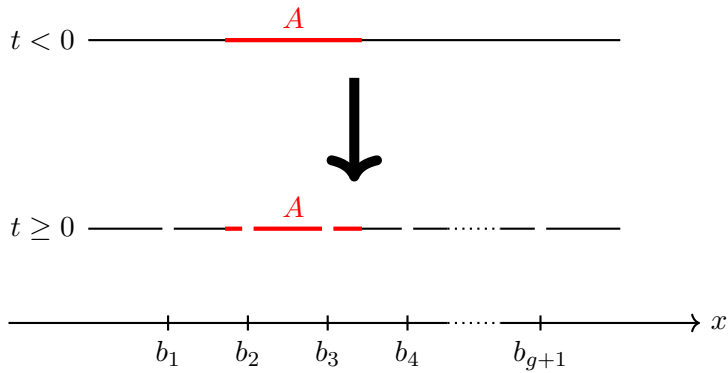
\begin{figure}[H]
\centering
\tikzmath{\l=3.5;\h=2.5;\b=0.75;\c=0.075;\a=1.5;\aa=3;\k=0.15;\kk=0.45;\f=0.45;\ff=1.5;\m=-2.25;\mm=1.5;\off=0.075;\p=0.3;}
\scalebox{1}{
\begin{tikzpicture}

    \def \b {-0.7*\l};
    \def \bb {-0.4*\l};
    \def \bbb {-0.1*\l};
    \def \bbbb {0.2*\l};
    \def \bbbbb {0.7*\l};
    
    \draw[thick] ($(-\l,0)$)node[left=1]{$t<0$} -- ($(\l,0)$);
    \draw[ultra thick, red] ($(\bb-\p, 0)$) -- ($(\bbb+1.5*\p, 0)$) node[pos=0.5, above]{$A$};
    
    \draw[thick] ($(-\l,-\h)$)node[left=1]{$t\geq0$} -- ($(0.35*\l,-\h)$);
    \draw[thick, dotted] ($(0.35*\l,-\h)$)-- ($(0.55*\l,-\h)$);
    \draw[thick] ($(0.55*\l,-\h)$) -- ($(\l,-\h)$);
    \draw[ultra thick, red] ($(\bb-\p, -\h)$) -- ($(\bbb+1.5*\p, -\h)$) node[pos=0.5, above]{$A$};
    
    \draw[->, line width=0.8ex] ($(0,-0.2*\h)$) -- ($(0,-0.75*\h)$);  

    \draw[line width=10pt, white] ($(\b-\c,-\h)$) -- ($(\b+\c,-\h)$);
    \draw[line width=10pt, white] ($(\bb-\c,-\h)$) -- ($(\bb+\c,-\h)$);
    \draw[line width=10pt, white] ($(\bbb-\c,-\h)$) -- ($(\bbb+\c,-\h)$);
    \draw[line width=10pt, white] ($(\bbbb-\c,-\h)$) -- ($(\bbbb+\c,-\h)$);
    \draw[line width=10pt, white] ($(\bbbbb-\c,-\h)$) -- ($(\bbbbb+\c,-\h)$);

    \draw[thick] ($(-1.3*\l,-1.5*\h)$)-- ($(0.35*\l,-1.5*\h)$);
    \draw[thick, dotted] ($(0.35*\l,-1.5*\h)$)-- ($(0.55*\l,-1.5*\h)$);
    \draw[thick, ->] ($(0.55*\l,-1.5*\h)$) -- ($(1.3*\l,-1.5*\h)$)node[right]{$x$};
    
    \draw[thick] ($(\b,-1.5*\h)+(0,-0.1)$)node[below]{$b_1$} -- ($(\b,-1.5*\h)+(0,0.1)$);
    \draw[thick] ($(\bb,-1.5*\h)+(0,-0.1)$)node[below]{$b_2$} -- ($(\bb,-1.5*\h)+(0,0.1)$);
    \draw[thick] ($(\bbb,-1.5*\h)+(0,-0.1)$)node[below]{$b_3$} -- ($(\bbb,-1.5*\h)+(0,0.1)$);
    \draw[thick] ($(\bbbb,-1.5*\h)+(0,-0.1)$)node[below]{$b_4$} -- ($(\bbbb,-1.5*\h)+(0,0.1)$);
    \draw[thick] ($(\bbbbb,-1.5*\h)+(0,-0.1)$)node[below]{$b_{g+1}$} -- ($(\bbbbb,-1.5*\h)+(0,0.1)$);
\end{tikzpicture}
}
\caption{Schematic visualization of the multi-splitting quench in 1+1D CFT. At time $t=0$, the infinite line is cut at positions $x=b_1, b_2,..., b_{g+1}$ yielding $N$ distinct spatial regions for times $t\geq0$.
}
\label{fig:setup_schematic}
\end{figure}

Local quenches -- local excitations of a vacuum state -- are a paradigmatic way of studying entanglement dynamics \cite{Calabrese_2016} and are realizable in ultra-cold atom systems (e.g. \cite{Kaufman_2016,Meinert_2013}). 
 Furthermore, two of the authors have argued that the entanglement dynamics of a strongly coupled system splitting into multiple subsystems has  relevance for the hadronization process in heavy-ion collisions \cite{Muller:2022htn}. We therefore apply our prescription to the calculation of the vacuum CFT being split into an arbitrary number of pieces.\footnote{In a heavy-ion collision the hadronization process can be understood as a splitting of a highly excited, quasi-thermal state into many pieces. In order to be directly applicable, our formalism will need to be extended to initial thermal states instead of the vacuum state. We will study this case in a forthcoming publication.} See figure \ref{fig:setup_schematic} for a schematic depiction.

The entanglement entropy of a subsystem $A$ as a function of time $S_A(t)$ can be calculated holographically using the Ryu-Takayanagi formalism by looking at the length of the smallest geodesic homologous to $A$ in the holographic dual to the worldsheet of the reduced density matrix $\rho_A(0)$ \cite{Ryu_2006}.
The appropriate worldsheet is as follows:
\begin{align}
    {\rm Tr}_{A^c}\braket{\psi''(x'')|\rho_A(t)|\psi'(x')}
    =Z^{-1}\braket{\psi''(x'')|e^{-itH}|0}\braket{0|e^{+itH}|\psi'(x')}
\end{align}
In order to render the path-integral absolutely convergent we introduce a regulator $a$ as \cite{Calabrese_2007}
\begin{align}
    {\rm Tr}_{A^c}\braket{\psi''(x'')|\rho_A(t)|\psi'(x')}
    =Z^{-1}\braket{\psi''(x'')|e^{-itH-aH}|0}\braket{0|e^{+itH-aH}|\psi'(x')}\label{eq:matrix_elements_splitting}\;.
\end{align}
We keep $a$ real and small, but non-zero, and take the limit $a\to 0$ only at the end.
Figure~\ref{fig:global_splitting_worldsheet} shows the appropriate worldsheet for calculating the matrix elements in eq.~\eqref{eq:matrix_elements_splitting} for the case of three cuts.

\begin{figure}[H]
\centering
\tikzmath{\taumax=5;\tmax=3;\l=1.2;\a=2.5;\ax=0.8;}
\tdplotsetmaincoords{70}{40}
\scalebox{0.8}{
\begin{tikzpicture}[tdplot_main_coords]
    \path[draw, pattern=hatch, pattern color=red!30!gray,hatch size=3pt] ($(-\l,\a,0)$) -- ($(-\l,\taumax,0)$) -- ($(\l,\taumax,0)$) -- ($(\l,\a,0)$) --cycle;
    \path[draw, fill=red!30!gray, semitransparent] ($(-\l,\a,0)$) -- ($(-\l,\a,\tmax)$) -- ($(\l,\a,\tmax)$) -- ($(\l,\a,0)$) --cycle;
    \path[draw, fill=red!30!gray, semitransparent] ($(-\l,\a,\tmax)$) -- ($(-\l,0,\tmax)$) -- ($(\l,0,\tmax)$) -- ($(\l,\a,\tmax)$) --cycle;
    \path[draw, fill=cyan!50!gray, semitransparent] ($(-\l,0,0)$) -- ($(-\l,-\a,0)$) -- ($(\l,-\a,0)$) -- ($(\l,0,0)$) --cycle;
    \node at ($(\l,-\a,0)$) [below right] {$-a-it$};
    \node at ($(\l,0,0)$) [below right] {$-it$};
    \node at ($(-\l,-\a,0)$) [above] {$|0\rangle$};
    \path[draw, fill=cyan!50!gray, semitransparent] ($(-\l,0,0)$) -- ($(-\l,0,\tmax)$) -- ($(\l,0,\tmax)$) -- ($(\l,0,0)$) --cycle;
    \path[draw, pattern=hatch, pattern color=cyan!30!gray,hatch size=3pt] ($(-\l,-\a,0)$) -- ($(-\l,-\taumax,0)$) -- ($(\l,-\taumax,0)$) -- ($(\l,-\a,0)$) --cycle;
    \draw[thick] ($(0,-\a,0)$)--($(0,0,0)$);
    \draw[thick] ($(0,\a,\tmax)$)--($(0,0,\tmax)$);
    \draw[thick] ($(0,0,0)$)--($(0,0,\tmax)$);
    \draw[thick] ($(0,\a,0)$)--($(0,\a,\tmax)$);

    \def \q {0.4*\l};
    \draw[thick] ($(-\q,-\a,0)$)--($(-\q,0,0)$);
    \draw[thick] ($(-\q,\a,\tmax)$)--($(-\q,0,\tmax)$);
    \draw[thick] ($(-\q,0,0)$)--($(-\q,0,\tmax)$);
    \draw[thick] ($(-\q,\a,0)$)--($(-\q,\a,\tmax)$);

    \draw[thick] ($(\q,-\a,0)$)--($(\q,0,0)$);
    \draw[thick] ($(\q,\a,\tmax)$)--($(\q,0,\tmax)$);
    \draw[thick] ($(\q,0,0)$)--($(\q,0,\tmax)$);
    \draw[thick] ($(\q,\a,0)$)--($(\q,\a,\tmax)$);
    \node at ($(-\l,0,\tmax)$) [left] {$0$};
    \node at ($(\l,\a,\tmax)$) [right] {$a$};
    \node at ($(\l,\a,0)$) [below right] {$a-it$};
    \node at ($(\l,\taumax,0)$) [below right] {$\infty-it$};
    \node at ($(\l,-\taumax,0)$) [below right] {$-\infty-it$};
    \node at ($(0,\a,0)$) [below] {$\langle0|$};

    \coordinate[] (o) at ($(-4*\l,0,0)$);
    \draw[thick, ->] (o) -- ($(o)+(\ax,0,0)$)node[right]{$x$};
    \draw[thick, ->] (o) -- ($(o)+(0,\ax,0)$)node[above]{$\tau$};
    \draw[thick, ->] (o) -- ($(o)+(0,0,\ax)$)node[left]{$t$};

    \draw[ultra thick, red] ($(-0.7*\l,0,\tmax)$)--($(0.3*\l,0,\tmax)$) node[pos=0.5, below]{$A$};
\end{tikzpicture}
}
\caption{The regulated worldsheet for calculating $\rho_A(t)$ of a three cut splitting quench where $i\tau=t$.
The infinite Euclidean time evolution, indicated by the checkered regions, serves as ground state preparation. The black lines from $-a$ to $a$ in the Euclidean time correspond to the cuts, i.e., the points which are excluded in the path integral.
}\label{fig:global_splitting_worldsheet}
\end{figure}
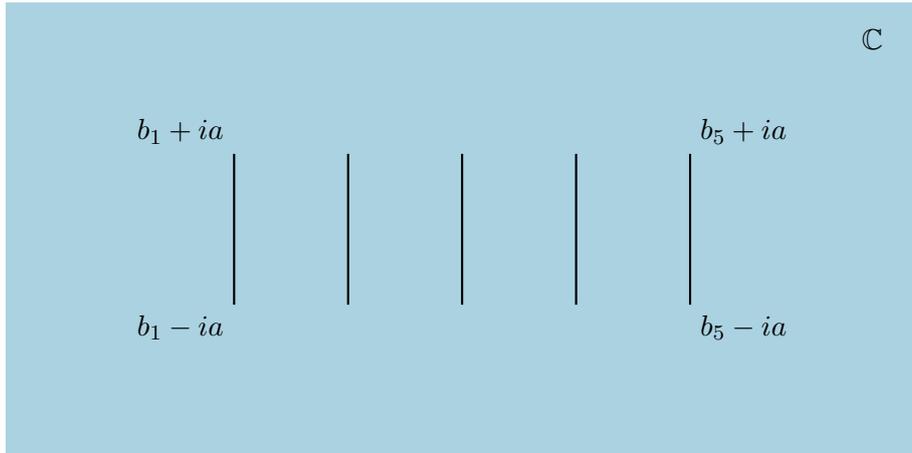
\begin{figure}[H]
\centering
\tikzmath{\l=3;\h=3;\b=1.5;\a=1;}
\scalebox{1}{
\begin{tikzpicture}

    \path[fill=cyan!50!gray, opacity=0.5] ($(-2*\l,-\h)$)--($(2*\l,-\h)$)--($(2*\l,\h)$)--($(-2*\l,\h)$)--cycle;

    \foreach \i in {-2,...,2}
    {
        \draw[thick] (\i*\b, -\a) -- (\i*\b, \a);
    }

    \path (-2*\b, \a) node[above left] {$b_1+ia$}
    -- (-2*\b, -\a) node[below left] {$b_1-ia$}
    -- (2*\b, \a) node[above right] {$b_5+ia$}
    -- (2*\b, -\a) node[below right] {$b_5-ia$}
    -- (1.8*\l,0.75*\h) node[above] {$\mathbb{C}$}
    ;

\end{tikzpicture}
}
\caption{The regulated worldsheet for calculating $\rho_A(0)$ of five cuts at positions $b_1,...,b_5$.}\label{fig:worldsheet_rho(0)}
\end{figure}
\begin{figure}[H]
\centering
\includegraphics[scale=.5]{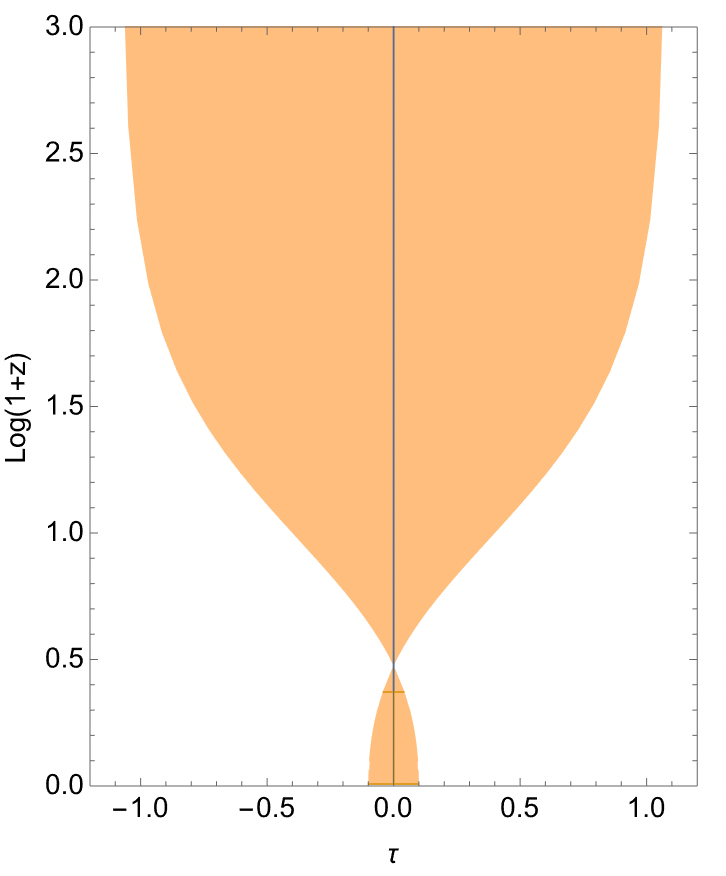}
\caption{An example of non-trivial behaviour of the holographic dual of a slit in the Euclidean worldsheet. This is the left-most slit (corresponding to the unit circle) of a multiple splitting quench at vanishing boundary tension ($T=0$) and regulator set to $a=0.1$. We project the surface onto the $(\tau,z)$ plane.}
\label{fig:brane}
\end{figure}

In order to recover the real-time dynamics of the quench, we  analytically continue the worldsheet for $\rho_A(0)$  from Euclidean to Lorentzian time\footnote{There is $\mathbb{Z}_2$ symmetry across the $t=0$ plane, which justifies the analytic continuation. See section 2 of \cite{Krasnov_2002} for a more detailed account.}.
Therefore, the appropriate worldsheet to consider the holographic dual of is the Euclidean plane missing $n$ line segments of length $2a$ at arbitrary placements along the real axis, as shown in figure~\ref{fig:worldsheet_rho(0)}.
This worldsheet is not simply connected and therefore has moduli. This renders this calculation intractable using the replica trick as the resulting Riemann surface is incredibly complicated. Using the Ryu-Takayanagi formalism without uniformizing the worldsheet is also intractable as the holographic duals of the line segments are not well-behaved and change dramatically as the regulator goes to 0 as discussed in section~\ref{sec:Holography_for_Multiple_Boundaries}. An example is plotted in Fig. \ref{fig:brane}.

After conformally mapping the worldsheet to its Schottky uniformization everything is well-behaved and one can take the $a\rightarrow0$ limit with no issues. The appropriate conformal map from the Schottky uniformization to our worldsheet is
\begin{align}
\begin{split}
    \tilde{\phi}(z)=&\;b_{g-1}\lb1-\frac{1}{1-z\alpha^{-1}}-\frac{
    1
    }{1-z\alpha}\rb
    \\\
    &+b_{g-1}\lb\sum_{i=1}^g q_i^2\frac{\alpha(z^2-1)(1-4\bar{d}_i\alpha +\alpha^2+\bar{d}_i^2(1+\alpha^2))}{(\alpha-\bar{d}_i)^2(\alpha \bar{d}_i-1)^2(\bar{d}_i-z)(\bar{d}_i z-1)}\rb+\mathcal{O}(
    \alpha^8)\;.
    \label{map}
\end{split}
\end{align}
See appendix \ref{sec:appendix_SKP_asymptotics} for the explicit derivation of this map (where here we specialize to real $\alpha$) from the infinite product representation of the Schottky-Klein prime function.

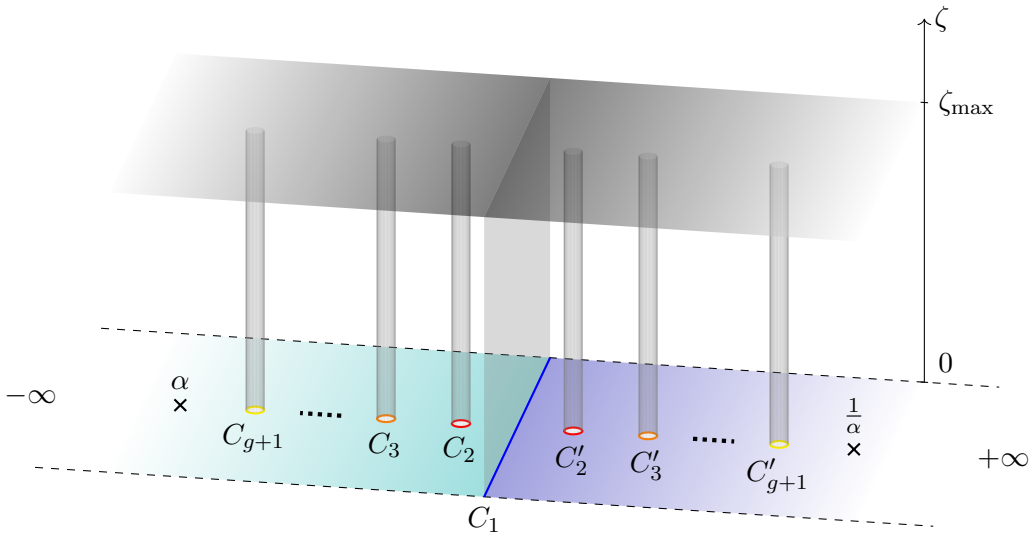
\begin{figure}[H]
\centering
\tikzmath{\l=5;\h=2.5;}
\tdplotsetmaincoords{68}{10}
\scalebox{1}{
\begin{tikzpicture}[tdplot_main_coords]
    \def \d {0.8*\l};
    \def \shang {65};

    \path[
    right color=cyan!50!gray, left color=white, opacity=0.5, shading angle=\shang
    ] (-\l, -\h, 0) --(0, -\h, 0) --(0, \h, 0) --(-\l, \h, 0) --cycle;
    \path[
    left color=blue!50!gray, right color=white, opacity=0.5, shading angle=\shang
    ] (\l,-\h,0)--(0,-\h,0)--(0,\h,0)--(\l,\h,0)--cycle;

    \draw[dashed] (-1.2*\l, -\h, 0)--(1.2*\l, -\h, 0);
    \draw[dashed] (-1.2*\l, \h, 0)--(1.2*\l, \h, 0);


    \path[fill=gray, opacity=0.3] (0,-\h,0)--(0,-\h,\d)--(0,\h,\d)--(0,\h,0)--cycle;
    \draw[thick, blue] (0,-\h,0)node[below]{$\color{black}C_1$}--(0,\h,0);

    
    \def \r {0.12};
    \coordinate (c2) at (-0.15*\l,0,0);
    \coordinate (c3) at (-0.35*\l,0,0);
    \coordinate (cg1) at (-0.7*\l,0,0);
    
    \path[draw=red, thick, fill=white]
    ($(c2) - (\r,0,0)$) arc(-180:180:\r)node[pos=0.25, below]{$C_2$}
    ;
    \path[draw=red, thick, fill=white]
    ($- 1*(c2)- (\r,0,0)$) arc(-180:180:\r)node[pos=0.25, below]{$C'_2$}
    ;

    \path[draw=orange, thick, fill=white]
    ($(c3) - (\r,0,0)$) arc(-180:180:\r)node[pos=0.25, below]{$C_3$}
    ;
    \path[draw=orange, thick, fill=white]
    ($- 1*(c3)- (\r,0,0)$) arc(-180:180:\r)node[pos=0.25, below]{$C'_3$}
    ;

    \path[draw=yellow, thick, fill=white]
    ($(cg1) - (\r,0,0)$) arc(-180:180:\r)node[pos=0.25, below]{$C_{g+1}$}
    ;
    \path[draw=yellow, thick, fill=white]
    ($- 1*(cg1)- (\r,0,0)$) arc(-180:180:\r)node[pos=0.25, below]{$C'_{g+1}$}
    ;

    \def \res {40};
    \foreach \i in {0,...,\res}
    {
        \draw[gray, opacity=0.2] ($- 1*(c2) - (360/\res*\i:\r)$)
        --($- 1*(c2) + (0,0,\d) - (360/\res*\i:\r)$)
        ;
        \draw[gray, opacity=0.2] ($(c2) - (360/\res*\i:\r)$)
        --($(c2) + (0,0,\d) - (360/\res*\i:\r)$)
        ;
        
        \draw[gray, opacity=0.2] ($- 1*(c3) - (360/\res*\i:\r)$)
        --($- 1*(c3) + (0,0,\d) - (360/\res*\i:\r)$)
        ;
        \draw[gray, opacity=0.2] ($(c3) - (360/\res*\i:\r)$)
        --($(c3) + (0,0,\d) - (360/\res*\i:\r)$)
        ;
        
        \draw[gray, opacity=0.2] ($- 1*(cg1) - (360/\res*\i:\r)$)
        --($- 1*(cg1) + (0,0,\d) - (360/\res*\i:\r)$)
        ;
        \draw[gray, opacity=0.2] ($(cg1) - (360/\res*\i:\r)$)
        --($(cg1) + (0,0,\d) - (360/\res*\i:\r)$)
        ;
    }

    \def \a {0.9*\l};
    \node [cross, label=above:{$\alpha$}] at (-\a,0,0) {};
    \node [cross, label=above:{$\frac{1}{\alpha}$}] at (\a,0,0) {};

    \path[
    left color=black, right color=white, opacity=0.5, shading angle=\shang
    ] (0, -\h, \d) --(\l, -\h, \d) --(\l, \h, \d) --(0, \h, \d);

    \path[
    right color=black, left color=white, opacity=0.5, shading angle=\shang
    ] (-\l, -\h, \d) --(0, -\h, \d) --(0, \h, \d) --(-\l, \h, \d);

    ;
    \draw[->] (1*\l,\h,0) -- (1*\l,\h, 1.3*\d) node[right]{$\zeta$};
    \draw[] (1*\l-0.05,\h,0)--(1*\l+0.05,\h,0) node[above right]{$0$};
    \draw[] (1*\l-0.05,\h,\d)--(1*\l+0.05,\h,\d) node[right]{$\zeta_{\rm max}$};

    \draw[dotted, ultra thick] ($(c3) - (5*\r,0,0)$)
    -- ($(cg1) + (5*\r,0,0)$);
    \draw[dotted, ultra thick] ($- 1*(c3) + (5*\r,0,0)$)
    -- ($- 1*(cg1) - (5*\r,0,0)$);

    \node[] at (-1.3*\l,0,0) {$-\infty$};
    \node[] at (1.3*\l,0,0) {$+\infty$};

\end{tikzpicture}
}
\caption{Holographic dual of the doubled identified half-strip associated with the Schottky uniformized worldsheet with $g+1$ cuts. The bulk boundaries of $C_i$ and $C'_i$ are identified for all $i\in\{2, ..., g+1\}$, respectively, as well as the dashed lines. The interior of the unit disk is mapped onto the half-strip which extends from $-\infty$ to $0$ (boundary $C_1$). Its double extends from $0$ to $+\infty$.}

\label{fig:bulk_extension}
\end{figure}

In order to have a globally defined metric and implement periodicities easier we map the unit disk to an identified half-strip with complex coordinates $\eta$ and $\bar{\eta}$, using the transformation
\be
\eta(z)=\frac{1}{2\pi}\log\lb z\rb\;,\label{eq:map_to_rectangle}
\ee
Figure~\ref{fig:bulk_extension} shows the doubled identified half-strip structure for the general case of $g+1$ cuts. The circles $C_2$, ..., $C_{g+1}$ in the unit disk are again sent to circles on the half-strip up to corrections of order $\mathcal{O}(a^2)$.

The Schwarzian derivative of the map from the $(\eta,\bar{\eta})$-space to the splitting quench worldsheet evaluates to
\be
L(\eta)=
\{\tilde{\phi}\lb z(\eta)\rb,\eta\}=\pi^2+\mathcal{O}(a^2)\;,
\ee
where $z(\eta)$ is the inverse of eq.~\eqref{eq:map_to_rectangle}.
As a result, the metric eq.~\eqref{eq:splitting_banados_metric} on the identified half-strip is given by the BTZ metric
\be 
ds^2=\frac{R^2}{\zeta^2}\lb\lb1-\frac{\pi^2\zeta^2}{2}\rb^2d\Xi^2+
\lb1+\frac{\pi^2\zeta^2}{2}\rb^2d\Upsilon^2+d\zeta^2\rb\label{eq:BTZ_metric}\;,
\ee
where $\eta=\Xi+i\Upsilon$. One can then solve for how the boundary extends into the bulk by enforcing the condition that $K\propto T$ \cite{Fujita_2011}, where $K$ is the trace of the extrinsic curvature on the brane and $T$ is the boundary tension. For $T=0$, everything extending straight into the bulk e.g. ($C_1$ extending as a vertical plane into the bulk, whereas $C_2$, ..., $C_{g+1}$ form cylinders) is a solution to the resulting differential equation. The resulting holographic dual geometry is shown schematically in figure~\ref{fig:bulk_extension}. We chose this value of $T$ as it drastically simplifies calculations and as stated earlier, alternative values only shift the disconnected geodesic lengths by a constant. As such we plot both geodesic contributions.



Given the holographic dual of the multi-splitting quench, it is now possible to calculate holographic entanglement entropy (HEE) of spatial subsystems.
According to the Ryu-Takayanagi prescription \cite{Ryu_2006}, HEE is determined by the area of the minimal codimension-two surface in the AdS bulk which shares the same endpoints as the subsystem.
For Riemann surfaces with boundaries, there exist two types of geodesics that do not violate the homology constraint: A connected one which links the endpoints without crossing any boundaries, and a disconnected one where each individual endpoint connects to the nearest boundary. In the case where the endpoints connect to different boundaries, a horizon which connects these boundaries must be added, resulting in an additional black hole entropy term.
The physical HEE of a spatial subsystem is then equal to the smallest of these geodesics.

Since the metric eq.~\eqref{eq:BTZ_metric} is the same as for the double splitting, we use the same geodesic expressions as derived in \cite{Caputa:2019avh,Lap:2024hsy}.
The individual parts of the disconnected geodesic are given by one half of the connected geodesic to the mirror image across the boundary.
In the following we apply the Schottky uniformization approach to the calculation of HEE after a splitting quench into four and more segments.
Instead of the entanglement entropy itself, which is divergent in the UV-cutoff $\epsilon$, it is useful to consider the entanglement entropy growth $\Delta S_A=S_A-S_A^{(0)}$, where $S_A^{(0)}=\frac{c}{3}\log\frac{l}{\epsilon}$ is the well known vacuum entropy for a subsystem of length $l$ in 1+1D CFT. This quantity is independent of the UV-cutoff of the underlying theory.\footnote{Geodesics in hyperbolic space are formally divergent as the metric diverges at the boundary. We have chosen the standard regulation procedure of taking the limit $\lim_{\epsilon\rightarrow0}$ of geodesics at $z=\epsilon>0$ and considering the difference with the vacuum result. In order to be more rigorous we could have constructed a one parameter family of regularizing surfaces as advocated in \cite{Krasnov:2000zq}}


\subsection{Three cuts (four segments)}

\begin{figure}[H]
\centering
\tikzmath{\l=3;\h=3;\r=4;}
\scalebox{1}{
\begin{tikzpicture}
    \def \s {0.15*\r};
    \def \t {0.1*\r};

    \path[fill=blue!50!gray, opacity=0.5] ($(-\l,-\h)$)--($(3*\l,-\h)$)--($(3*\l,\h)$)--($(-\l,\h)$)--cycle;
    \draw (0,0) node[circle, fill=white, minimum size=\r cm]{};
    
    \draw (0,0) node[circle, fill=cyan!50!gray, fill opacity=0.5, minimum size=\r cm, draw=blue, thick](c0) {};

    \draw[thick, orange] (0.3*\r, 0) -- (0.5*\r, 0);
    \draw[thick, orange, dashed] (0.3*\l, 0) -- (4.5/4*\r,0);

    \draw[thick, black] (0.125*\r,0) -- (0.3*\r,0);
    \draw[thick, black, dashed] (4.5/4*\r,0) -- (7/4*\r,0);

    \draw[thick, yellow] (0.02*\r, 0) -- (0.125*\r, 0);
    \draw[thick, yellow, dashed] (7/4*\r, 0) -- (2.8*\l, 0);
    
    \draw[thick, purple] (0.02*\r, 0) -- (-0.5*\r, 0);
    \draw[thick, purple, dashed] (-0.5*\r, 0) -- (-\l, 0);
    \draw[thick, purple, dashed] (2.8*\l, 0) -- (3*\l, 0);
    
    \draw ($(0.3*\r,0)$) node[circle, fill=white, minimum size=\s cm, draw=green, thick](c1) {};
    \draw ($(0.125*\r,0)$) node[circle, fill=white, minimum size=\t cm, draw=red, thick](c2) {};

    \def \ss {2*\s};
    \def \tt {4.5*\t};
    \draw ($(4.5/4*\r,0)$) node[circle, minimum size=\ss cm, draw, green, thick, fill=white](cp1) {};
    \draw ($(7/4*\r,0)$) node[circle, minimum size=\tt cm, draw, red, thick, fill=white](cp2) {};

    \draw[thick, <-] (cp1.south west) to [out=-150,in=-30] node[pos=0.5, below]{$\theta_2(z)$} (c1.south east);
    \draw[thick, <-] (cp2.north west) to [out=150,in=60] node[pos=0.5, above]{$\theta_3(z)$} (c2.north);

    \path (c0.south) node[below] {$C_1$}
    -- (c1.south) node[below] {$C_2$}
    -- (cp1.south) node[below] {$C'_2$}
    -- (c2.south) node[below] {$C_3$}
    -- (cp2.south) node[below] {$C'_3$}
    -- (2.8*\l,0.75*\h) node[above] {$\mathbb{C}$}
    ;

    \node [cross, label=above:{$\alpha$}] at (0.02*\r,0) {};
    \node [cross, label=above:{$\frac{1}{\alpha}$}] at (2.8*\l,0) {};

    \node[] at (-0.2*\r,0.2*\r) {$F$};
    
\end{tikzpicture}
}
\caption{Schottky uniformization of the genus 2 surface (the first case where the generators of the Schottky group do not commute). The $\theta_i(z)$ send interior circles to their mirror images across the unit circle $C_1$.
}\label{Schottky3}
\end{figure}
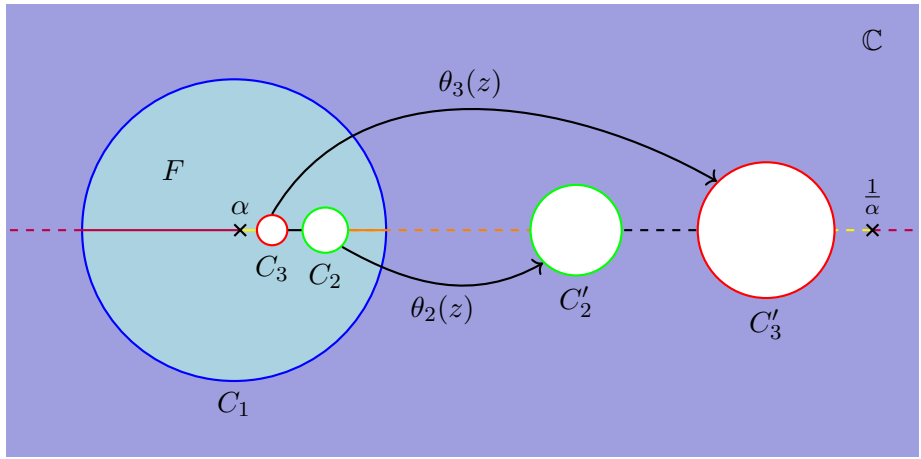

A splitting quench with three cuts is the first non-trivial check of our formalism. The topology of the world-sheet is that of a genus 0  Riemann surface with three boundaries (``pair of pants'') and therefore its Schottky uniformization is the unit disk with two circles excised as depicted in figure~\ref{Schottky3}.
Since there are two interior circles, we have two non-commuting generators of the Schottky group and the Schottky-Klein prime function cannot be written in a closed form. However, using the asymptotics for small $a$ derived in appendix~\ref{sec:appendix_SKP_asymptotics}, we arrive at the approximate conformal map eq.~\eqref{map} with $g=2$.

Holographic computations are then performed on the identified rectangle geometry, shown in figure~\ref{fig:bulk_extension}.
As explained in section~\ref{sec:BCFT_and_splitting_quenches}, HEE is determined by the connected and disconnected geodesic lengths between the endpoints of the spatial subsystem.

\begin{figure}[t]
\centering
\subfloat[subsystems\label{fig:3_cuts_subsystems}]{%
  \includegraphics[width=0.45\linewidth]{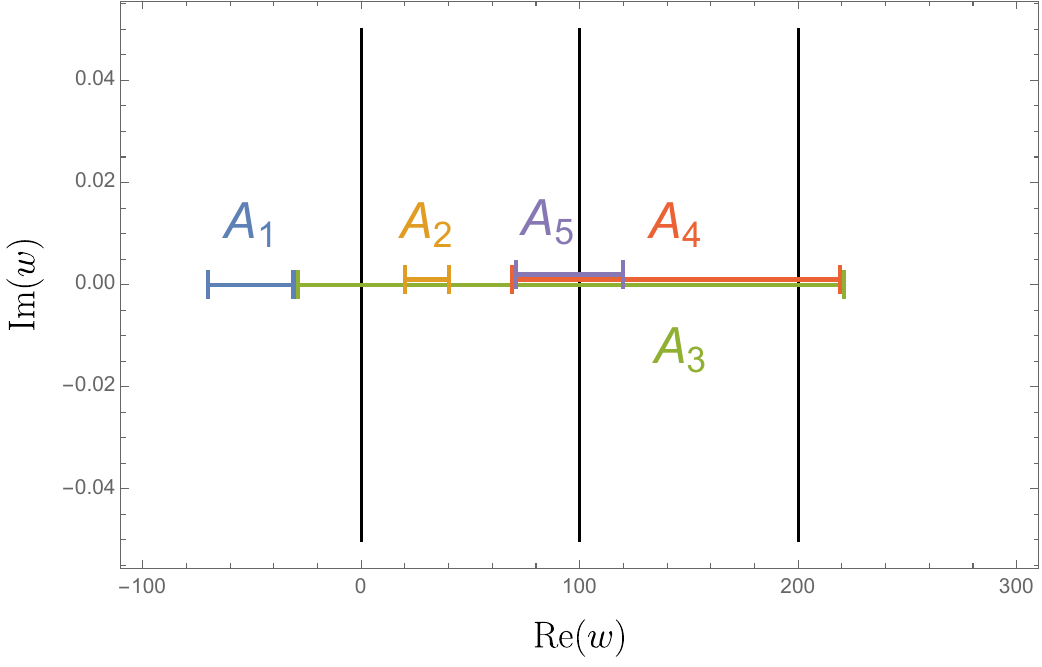}%
}
\quad
\subfloat[$A_1=[-70,-30{]}$\label{fig:3_cuts_-70_-30}]{%
  \includegraphics[width=0.45\linewidth]{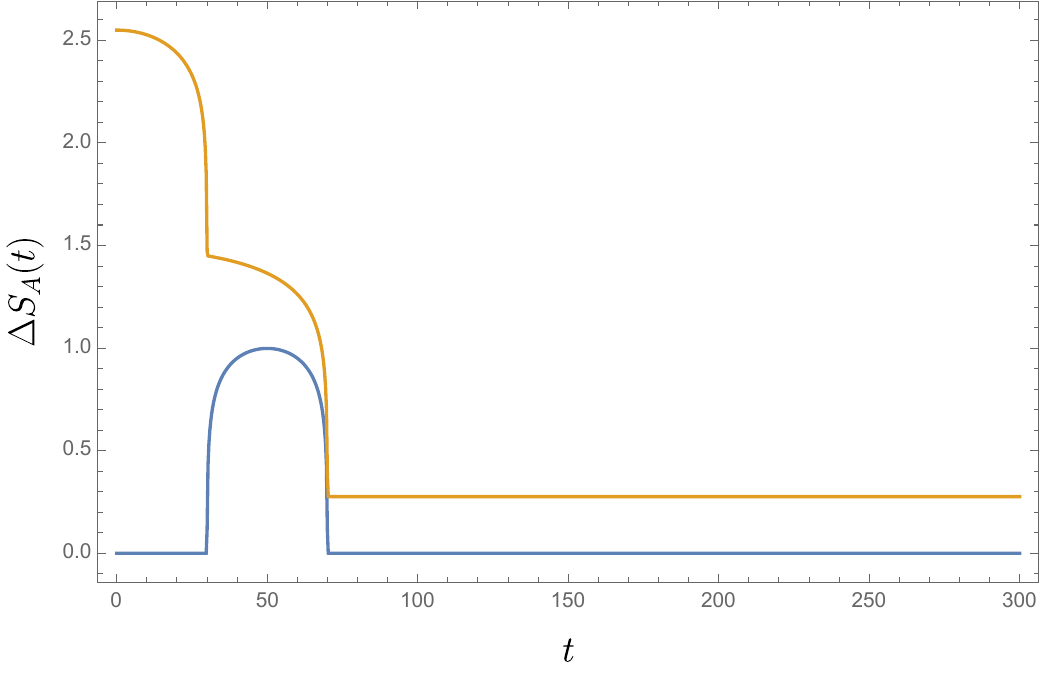}%
}
\\
\subfloat[$A_2=[20,40{]}$\label{fig:3_cuts_20_40}]{%
  \includegraphics[width=0.45\linewidth]{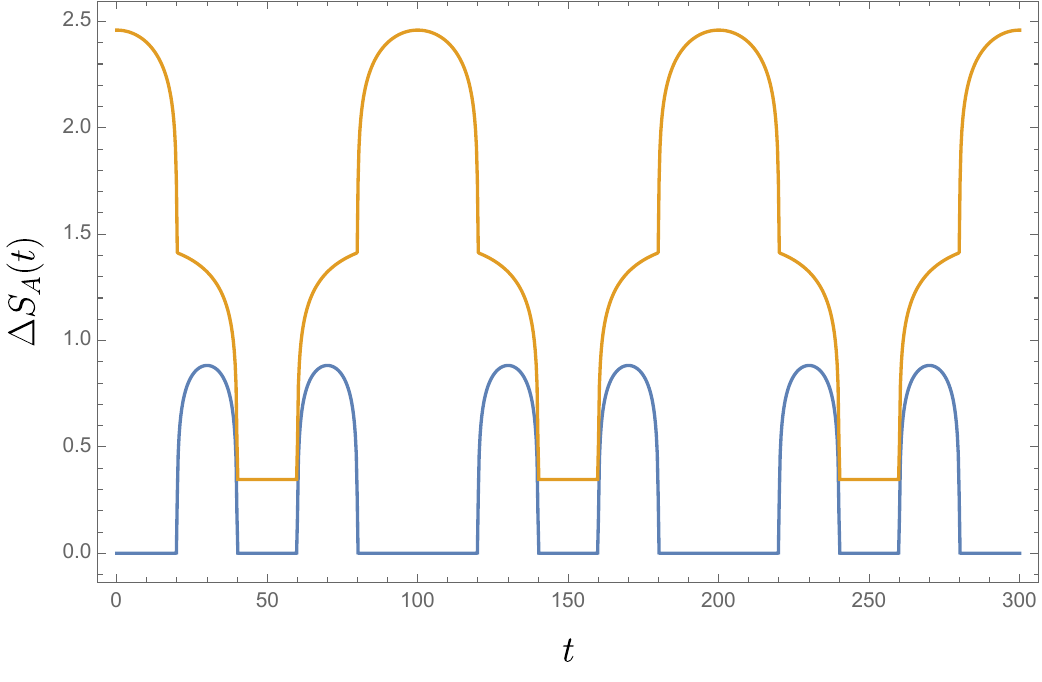}%
}\quad
\subfloat[$A_3=[-30,220{]}$\label{fig:3_cuts_-30_220}]{%
  \includegraphics[width=0.45\linewidth]{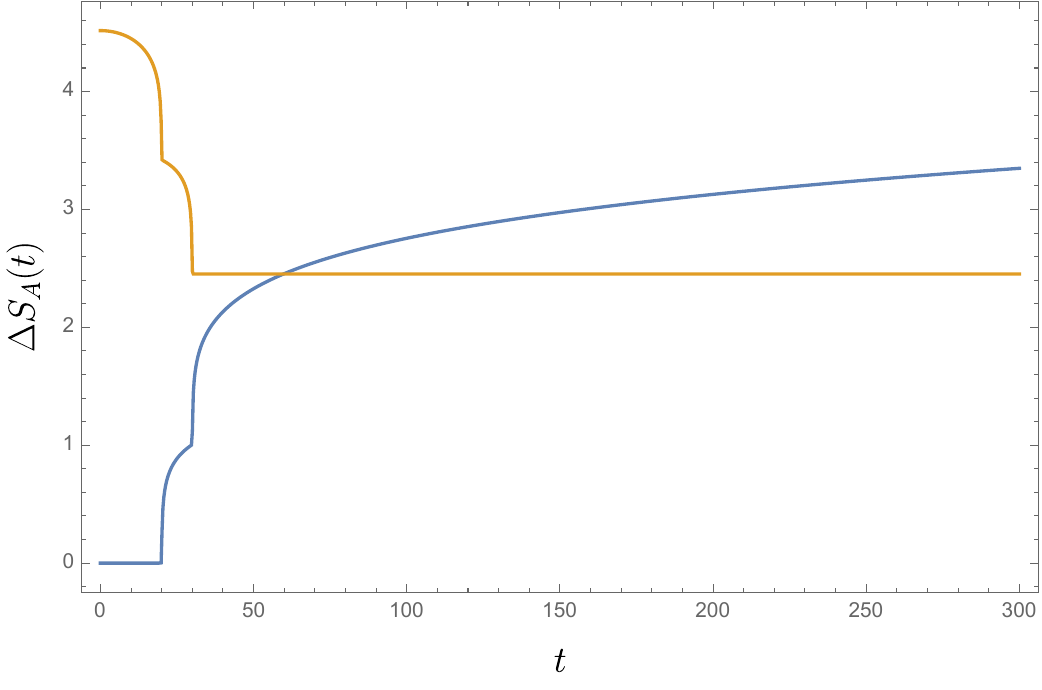}%
}
\\
\subfloat[$A_4=[70,220{]}$\label{fig:3_cuts_70_220}]{%
  \includegraphics[width=0.45\linewidth]{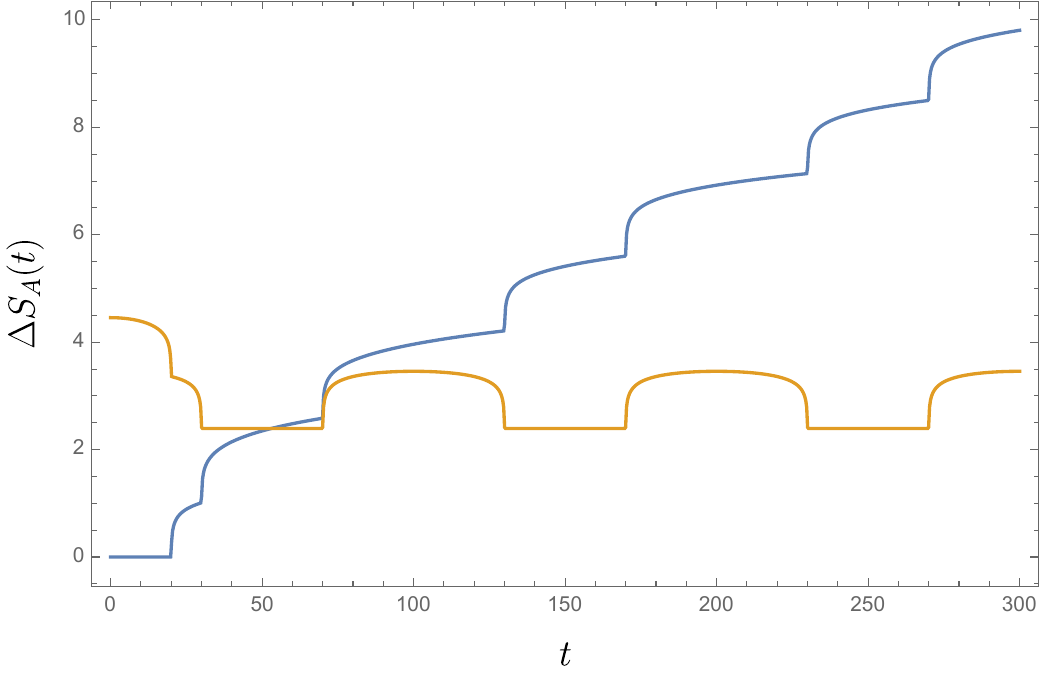}%
}\quad
\subfloat[$A_5=[70,120{]}$\label{fig:3_cuts_70_120}]{%
  \includegraphics[width=0.45\linewidth]{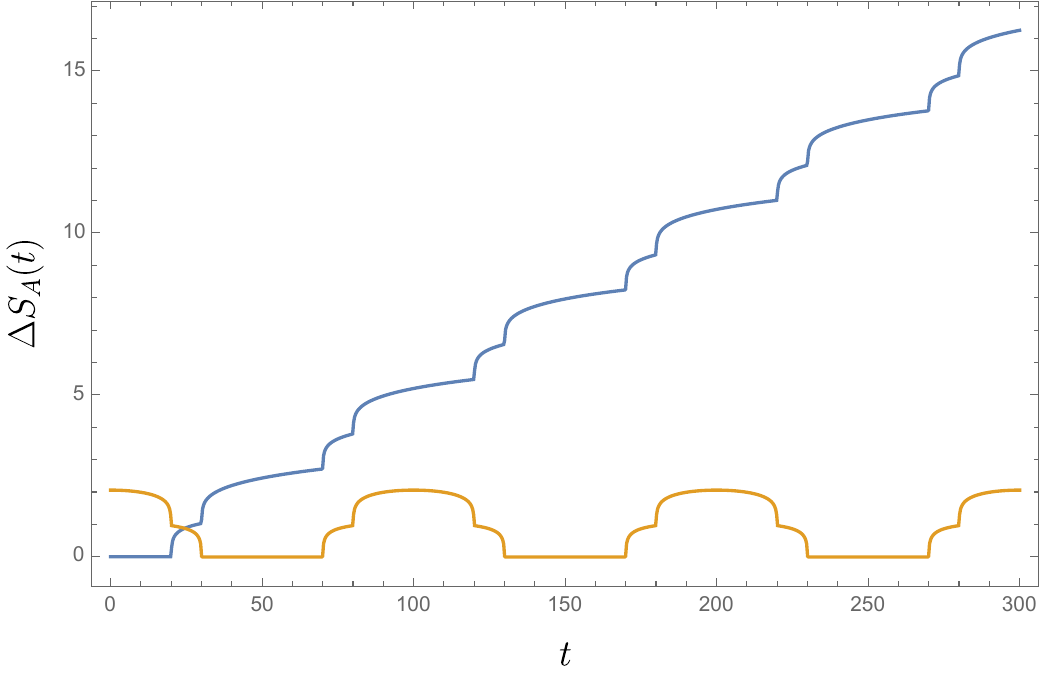}%
}
\caption{
Holographic entanglement entropy growth $\Delta S_A(t)$ (normalized by central charge) for a system split at $b_1=0$, $b_2=100$, and $b_3=200$ with boundary tension $T=0$.
(a) Visualization of the cuts and subsystems.
For better visibility some subsystems are shifted upwards, but all are defined to be on the real axis. Also, we mark the endpoints by vertical lines.
(b)--(f) Real time evolution of $\Delta S_A(t)$ for various subsystems, $A=A_1, ..., A_5$. The blue and orange graphs correspond to the connected and disconnected contribution respectively.}

\label{fig:3_cuts}
\end{figure}

The true entanglement entropy is the minimum of the connected and disconnected geodesic contributions, i.e., the blue and orange lines in the time evolution plots of figure~\ref{fig:3_cuts}. However, since the boundary entropy $T$, which depends on the chosen conformally invariant boundary conditions, shifts the disconnected contribution vertically by a constant, we show both the disconnected and connected geodesic contributions separately. 

The novelty arising in the genus 2 case is that one can now have subsystems whose endpoints lie on two different finite segments (e.g., $A_5$ in figure~\ref{fig:3_cuts_subsystems}). In the quasiparticle picture\footnote{It is common to consider massless quasi-particles formed at the site of the local quench that then propagate with light speed outward and reflect off of boundaries. This gives good intuition as to the timescales on which things occur e.g. changes in the entanglement entropy.} we expect this to be significantly different as the excitations inside the line segment cannot escape out of the subsystem. Compare figure~\ref{fig:3_cuts_-30_220}, which quickly reaches equilibrium, with figure~\ref{fig:3_cuts_70_220} , which continues to periodically oscillate around its equilibrium value. While the connected geodesic goes off to infinity (avoiding the ``end-of-the-world brane") the disconnected geodesic becomes the minimum, and therefore the entanglement entropy behaves qualitatively similar to the behavior seen in figure~\ref{fig:3_cuts_70_220}.

\subsection{Many cuts}

We now argue that nothing novel occurs for more than three cuts. The endpoints of the entangled subsystem can either be in (1) the same finite region, (2) two different finite regions, (3) a finite region and an infinite region, (4) the two infinite regions. We give an example for arbitrarily chosen $N=16$, but the formalism works for all $N$
including $N\rightarrow\infty$.
\begin{figure}[H]
\centering
\hfill
\includegraphics[width=0.45\linewidth]{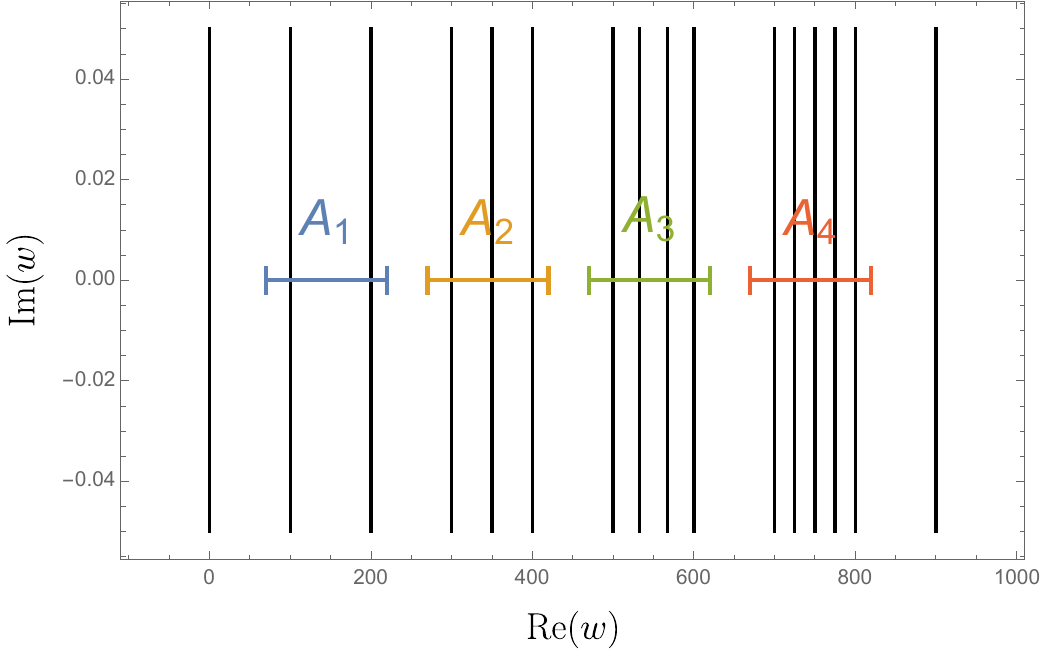}
\hfill
\includegraphics[width=0.45\linewidth]{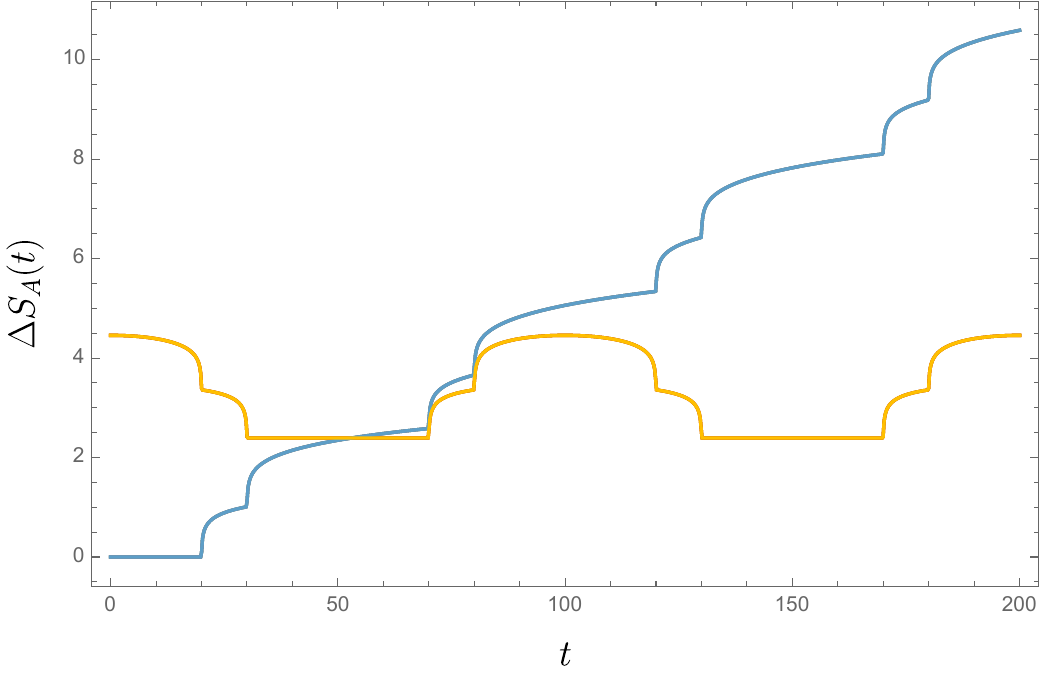}
\hfill
\caption{
Holographic entanglement entropy growth $\Delta S_A(t)$ (normalized by central charge) for a system split at $\{b_1, ..., b_{16}\}=\{0, 100, 200, 300, 350, 400, 500, 533.3, 566.7, 600, 700, 725, 750, 775, 800, 900\}$ with boundary tension $T=0$. The subsystems $A_1=[70,220]$, $A_2=[270,420]$, $A_3=[470,620]$, $A_4=[670,820]$
are chosen such that they are equivalent as seen from the endpoints and only differ in the number of internal cuts.
Left: Visualization of the cuts and subsystems.
Right: Real time evolution of $\Delta S_A(t)$ for various subsystems, $A_1$, $A_2$, $A_3$, and $A_4$. The blue and orange graphs correspond to the connected and disconnected contribution respectively.
The curves for the different subsystems are identical.}
\label{fig:multisplit}
\end{figure}

As one can see nothing novel occurs with more than three cuts. Since the entanglement entropy only depends on the length and endpoints of the subsystem it is blind to any structure interior to the cuts closest to its endpoints. In Fig. \ref{fig:multisplit} we choose 4 subsystems of length $150$, with 2, 3, 4, and 5 boundaries between the endpoints, respectively. With the same boundary conditions these all give the same entanglement entropy. We can make sense of this in the quasi-particle picture as the quasi-particles generated by the splits inside an interior segment will reflect on the internal boundaries and never escape from the subsystem. Therefore, the entanglement entropy never ``knows" about the interior boundaries. 

\section{Conclusion}

We introduced a formalism for dealing with boundary conformal field theories with multiple boundaries. Any worldsheet with two or more boundaries has moduli. Uniformizing the worldsheet allows one to better deal with the moduli, holographically or otherwise. In the limit that all moduli are small we give novel asymptotics for the Schottky-Klein prime function on the worldsheet which allows us to explicitly construct a conformal map to the Schottky uniformization of the worldsheet. As a proof of concept we give a novel calculation of holographic entanglement entropy for 1+1D CFTs on the real line which split into $N\geq 3$ segments. We show that for more than three splits the entanglement entropy is insensitive to boundaries interior to the outermost ones, and therefore nothing qualitatively changes for $N\geq4$.

Our results are applicable to any BCFT with multiple boundaries and the asymptotics to any Riemann surface with small moduli. The results for holographic entanglement entropy cannot be matched with non-holographic replica trick calculations as they are intractable for multiple boundaries. However, these dynamics are within experimental reach. AdS/BCFT has been extended to the Ising model\footnote{This is notable as the traditional lore is that only CFTs with large central charge (as opposed to the $c=1/2$ Ising model) are holographic.} \cite{Karch_2021} which is experimentally realizable on 1+1D spin chains tuned to criticality \cite{sahay2024emergentholographicforcestensor}. Such systems are part of the original motivation of studying splitting quenches \cite{Greiner_2002}. It would be interesting to see if multiple splitting quenches are realizable and reproduce our calculations.
We plan to apply this formalism to the calculation of additional observables (not just entanglement entropy) and to other worldsheets of interest, such as thermal states.

\acknowledgments
JDL would like to acknowledge the Belgian American Educational Foundation for its support. JDL and BM were supported by a research grant (DE-FG02-05ER41367) from the U.S. Department of Energy Office of Science. BM also acknowledges support from Yale University during extended visits in 2022 and 2023. CS acknowledges support from Duke University during a visit in 2023.

\appendix

\section{Proof of asymptotics for the Schottky-Klein prime function}\label{sec:appendix_SKP_asymptotics}
In \cite{Lap:2024hsy} we discussed the case of a Riemann surface with two boundaries. The associated Schottky group has only one generator (the M\"obius map from the inner to the outer circle) and therefore a closed form of the Schottky-Klein prime function is easily written down. For the case of three or more boundaries we have multiple generators which do not commute, and the infinite product becomes unwieldy. Thankfully, in our case the moduli are small parameters and therefore higher order terms become subleading contributions. We give a proof by induction of why we only need to consider the first level and that the constant contributions cancel.
A generic M\"obius transformation that sends a circle in the unit disk to its reflection across the unit disk (see figure~\ref{Schottky3}) is written as:
\be
\theta_i(z)=d_i+\frac{z}{1-z\bar{d}_i}q_i^2,
\label{Mobius}
\ee
where $d_i$ is the position of the center of the circle and $q_i$ is its radius. A level-two element of the Schottky group then has the form:
\[
\theta_j(\theta_i(z))=d_j+\frac{q_j^2(d_i+q_i^2z-d_i\bar{d}_iz)}{1-(d_i+q_i^2z)\bar{d}_j+z\bar{d}_i(d_i\bar{d}_j-1)},
\]
which in the limit $q_i,q_j\to 0$ becomes:
\[
\theta_j(\theta_i(z))\sim d_j+\frac{d_iq_j^2}{1-d_i\bar{d}_j}+\frac{zq_i^2q_j^2}{(1-z\bar{d}_i)(1-d_i\bar{d}_j)^2}+\mathcal{O}(q^6)
\]
Now assume that compositions of level $n$ are of the form:
\be 
f_{n}(x)\sim \sum_{i=0}^{n-1} g_i(\vec{d})q^{2i}+h(x)q^{2n}+\mathcal{O}(q^{2n+2}),
\label{mobcomp}
\ee
where there is no $x$ dependence until the $q^{2n}$ term and we are using the shorthand notations $\vec{d}$ for arguments depending on all $d_i$ and $q^{2n}$ for products of $q_i$ such that the total power is $2n$. Since $\partial_x f_n(x)=\mathcal{O}(q^{2n})$, the chain rule tells us that:
\[\partial_x \theta_k(f_n(x))=\frac{q_k^2}{(1-f_n(x)\bar{d}_k)^2}\partial_xf_n(x)=\mathcal{O}(q^{2n+2})\]

Explicitly we have:
\[
\theta_k(f_n(x))\sim d_k+\frac{g_0(\vec{d})}{1-\bar{d}_kg_0(\vec{d})}q_k^2+...+h'(x) q^{2n+2}+\mathcal{O}(q^{2n+4})
\]
which is of the necessary form:
\[
\theta_k(f_n(x))\sim\sum_{i=0}^{n}g'_i(\vec{d})q^{2i}+h'(x)q^{2n+2}(x)+\mathcal{O}(q^{2n+4})=f_{n+1}(x).
\]
Summarizing, we have shown that M\"obius map compositions of level 1 and 2 have this form and that, if a level $n$ composition has this form, then a level $n+1$ composition will, too. Thus, by induction, M\"obius map compositions of any level have the form (\ref{mobcomp}).

Now let us consider the Schottky-Klein prime function with this in mind. Remember that the Schottky-Klein prime function has an infinite product representation due to Baker \cite{baker_abelian_nodate}:
\be 
\omega(z,\alpha)=(z-\alpha)\prod_{\theta_k\in\Tilde{\Theta}}\frac{( z-\theta_k(\alpha))(\alpha-\theta_k(z))}{(\alpha-\theta_k(\alpha))(z-\theta_k(z))}\;,
\ee
where the product is over elements of the Schottky group excluding the identity and inverses and $\alpha$ and $z$ are points in the domain $\Omega$ (the disconnected region formed by the interior of the unit disc with $n$ discs missing).
We then take the logarithm of the Schottky-Klein prime function to find asymptotics for the conformal map:
\[\ln(\omega(z;
\alpha))=\ln(z-\alpha)+\sum_{\theta_k}\ln(\theta_k(z)-\alpha)+\ln(\theta_k(\alpha)-z)-\ln(\theta_k(z)-z)-\ln(\theta_k(\alpha)-\alpha)\]
Many cancellations occur in this expression if we take the $q_i\to 0$ limit. Consider
\[
\ln(f_n(a)-b)-\ln(f_n(b)-b)\;.
\]
All terms of less than leading order cancel as they do not depend on the argument of the function:
\[
\ln(f_n(a)-b)-\ln(f_n(b)-b)\propto q^{2n}+\mathcal{O}(q^{2n+2})
\]
Since we only need the leading order in $q$ -- remember that all $q_i$ are functions of the regulator which is sent to 0 at the end of the calculation -- it is sufficient to truncate to the level 1 elements of the Schottky group:
\begin{align}
\begin{split}
\ln(\omega(z;
\alpha))\sim\ln(z-\alpha)+\sum_{i=1}^g &\left[\ln(\theta_i(z)-\alpha)+\ln(\theta_i(\alpha)-z)\right.
\\\
&\left.\;-\ln(\theta_i(z)-z)-\ln(\theta_i(\alpha)-\alpha)\right]+\mathcal{O}(q^4)
\end{split}
\end{align}
Inserting our expression (\ref{Mobius}) for a generic M\"obius map yields:
\begin{align}
\begin{split}
\ln(\omega(z;
\alpha))\sim\ln(z-\alpha)
+\sum_{i=1}^g &\left[\ln(d_i+\frac{q_i^2z}{1-z\bar{d}_i}-\alpha) +\ln(d_i+\frac{q_i^2\alpha}{1-\alpha\bar{d}_i}-z) \right.
\\
&\left.\; -\ln(d_i+\frac{q_i^2z}{1-z\bar{d}_i}-z)-\ln(d_i+\frac{q_i^2\alpha}{1-\alpha\bar{d}_i}-\alpha)\right].
\end{split}
\end{align}
The conformal map for $n=g+1$ parallel slits perpendiciular to the $x$-axis is \cite{crowdy_solving_2020}:
\[
\phi(z)=\frac{1}{2\bar{\alpha}}+\frac{1}{2\alpha}-\partial_\alpha \ln(\omega(z;\alpha))-\frac{1}{\bar{\alpha}^2}\partial_{\bar{\alpha}^{-1}}\ln(\omega(z;\bar{\alpha}^{-1}))\;,
\]
Therefore, the approximate asymptotic conformal map is:
\begin{align}
\begin{split}
\phi(z)=\frac{1}{2\bar{\alpha}}&+\frac{1}{2\alpha}
-\frac{1}{\alpha-z}-\frac{1}{\bar{\alpha}^2(\bar{\alpha}^{-1}-z)}
\\
&+\sum_{i=1}^g \left[ q_i^2\frac{(\alpha-z)(z+\alpha(1+\bar{d}_i^2-2\bar{d}_i z)+\bar{d}_i(\bar{d}_i z-2))}{(\alpha-\bar{d}_i)^2(\alpha \bar{d}_i-1)^2(\bar{d}_i-z)(\bar{d}_i z-1)}\right. 
\\
&\left.\hspace{1.2cm}+q_i^2\frac{(\bar{\alpha}^{-1}-z)(z+\bar{\alpha}^{-1}(1+\bar{d}_i^2-2\bar{d}_i z)+\bar{d}_i(\bar{d}_i z-2))}{(\bar{\alpha}^{-1}-\bar{d}_i)^2(\bar{\alpha}^{-1} \bar{d}_i-1)^2(\bar{d}_i-z)(\bar{d}_i z-1)}\right] .
\end{split}
\end{align}

\subsection{The Map}

We repeat the procedure of \cite{Lap:2024hsy} and solve for the $q_i$ and $d_i$ as a function of the slit/regulator length $a$ and the placement of the slits $b_1,...,b_{g+1}$ perturbatively. We set the origin at $b_1=0$ for convenience and choose the pre-image of infinity to be $\alpha=\frac{a}{2b_{g+1}}$ so it goes to the origin nicely in the limit $a\to 0$. This yields:
\[
d_i=\frac{a(b_i+1)}{2b_ib_{g+1}}+\mathcal{O}(a^3)
\]
\[
d_{g-1}=\frac{a}{b_{g+1}}+\mathcal{O}(a^3)
\]
\[
q_i=\frac{a^2}{4b^2_{g+1}b^2_i}+\mathcal{O}(a^3)
\]
\[
q_{g-1}=\frac{a^2}{4b^2_{g+1}}+\mathcal{O}(a^3)
\]
The asymptotic conformal map with the appropriate prefactor is then:
\begin{align}
\begin{split}
\tilde{\phi}(z)=\alpha b_{g+1}\phi(z) =\, & a\frac{1-z^2}{2z} +a^2\frac{1-z^4}{4b_{g+1}z^2}
\\
& +a^3\frac{1}{8b_i^2z^2} \left(\frac{b_{g+1}^2z^2}{(b_i-b_{g+1})^2}-z^4+\frac{b_i^2(1-z^2)(1+z^2)^2}{b_{g+1}^2}\right) +\mathcal{O}(a^4)
\end{split}
\end{align}

\bibliographystyle{JHEP}
\bibliography{main.bib}

\end{document}